\documentclass[fleqn,usenatbib]{mnras}

\usepackage{newtxtext,newtxmath}

\usepackage[T1]{fontenc}

\DeclareRobustCommand{\VAN}[3]{#2}
\let\VANthebibliography\thebibliography
\def\thebibliography{\DeclareRobustCommand{\VAN}[3]{##3}\VANthebibliography}

\usepackage{graphicx}	
\usepackage{amsmath}	
\usepackage{amssymb}	

\newcommand{\dm}{pc\,cm$^{-3}$}

\newcommand{\us}{\textrm{$\mu$}s}

\title[FRB20191107B, and the origins of scattering]{The ultra narrow FRB20191107B, and the origins of FRB scattering}

\author[Gupta, V. et al.]{\parbox{\textwidth}
   {V.\ Gupta$^{1}\thanks{Corresponding Author: vivek.gupta@csiro.au}$, 
    C.\ Flynn$^{1,2}$,
    W.\ Farah$^{1, 3}$,
    M.\ Bailes$^{1,2}$,
    A.\ T.\ Deller$^{1, 2}$,
    C.\ K.\ Day$^{1, 4}$,
    M.\ E.\ Lower$^{1, 4}$
    }  \\ \\ \\
\parbox{\textwidth}{
$^{1}$Centre for Astrophysics and Supercomputing,
  Swinburne University of Technology, Mail H30, PO Box 218, VIC 3122,
  Australia\\
$^{2}$OzGrav: ARC Centre of Excellence for Gravitational Wave Discovery, Hawthorn, VIC 3122, Australia\\
$^{3}$SETI Institute 189 Bernardo Ave, Suite 200 Mountain View, CA 94043, United States\\
$^{4}$CSIRO, Astronomy and Space Science, Australia Telescope National Facility, PO Box 76, Epping, NSW 1710, Australia\\
}
}

\date{Accepted 2022 June 6. Received 2022 May 26; in original form 2021 October 1}

\pubyear{2022}

\begin{document}
\label{firstpage}
\pagerange{\pageref{firstpage}--\pageref{lastpage}}
\maketitle

\begin{abstract}

We report the detection of FRB20191107B with the UTMOST radio telescope at a dispersion measure (DM) of 714.9 ${\rm pc~cm^{-3}}$. The burst consists of three components, the brightest of which has an intrinsic width of only 11.3 $\mu$s and a scattering tail with an exponentially decaying time-scale of 21.4 $\mu$s measured at 835 MHz. We model the sensitivity of UTMOST and other major FRB surveys to such narrow events. We find that $>60\%$ of FRBs like FRB20191107B are being missed, and that a significant population of very narrow FRBs probably exists and remains underrepresented in these surveys. The high DM and small scattering timescale of FRB20191107B allows us to place an upper limit on the strength of turbulence in the Intergalactic Medium (IGM), quantified as scattering measure (SM), of ${\rm SM_{IGM} < 8.4 \times 10^{-7} ~kpc~m^{-20/3}}$. Almost all UTMOST FRBs have full phase information due to real-time voltage capture which provides us with the largest sample of coherently dedispersed single burst FRBs. Our 10.24 $\mu$s time resolution data yields accurately measured FRB scattering timescales. We combine the UTMOST FRBs with 10 FRBs from the literature and find no obvious evidence for a DM-scattering relation, suggesting that IGM is not the dominant source of scattering in FRBs. We support the results of previous studies and identify the local environment of the source in the host galaxy as the most likely region which dominates the observed scattering of our FRBs.

\end{abstract}

\begin{keywords}
(transients:) fast radio bursts -- scattering -- (cosmology:) large-scale structure of Universe 
\end{keywords}



\section{Introduction}

Fast Radio Bursts (FRBs) are very short duration ($\mu$s - ms) bursts of extragalactic origin and observed in radio frequencies. The number of detected FRBs has been growing steadily since their discovery in 2007 \citep{Lorimer2007} and much more rapidly over the last couple of years largely due to the commissioning of multiple wide Field-of-View (FoV) telescopes like the Australian SKA Pathfinder telescope (ASKAP; \citealt{Shannon2018, Macquart2020_IGM_Baryons_DM_z_relation}) and the Canadian Hydrogen Intensity Mapping Experiment (CHIME; \citealt{First_CHIME_catalog_2021}) telescope with dedicated FRB search surveys. FRBs show a characteristic dispersion sweep in their dynamic spectra caused by the frequency dependent delay induced by propagation through ionised plasma along their line of sight. Dispersion Measure (DM) is used to quantify this dispersion sweep and is proportional to the path integral of the electron density along the propagation path:
\begin{equation}
    {\rm DM} = \int_0^z \frac{n_e(z')}{(1 + z')}dl,
\end{equation}
where $n_e(z')$ is the physical electron density at redshift $z'$. 
Most FRBs have total observed DMs much larger than the predicted Milky Way contribution derived from the electron density models of the ionised Interstellar Medium (ISM) in our galaxy \citep{NE2001, YMW16}, with the exception of the galactic magnetar SGR1935$+$2154 which exhibited FRB like emission in April 2020 \citep{Bochenek2020_STARE2_FRB, CHIME2020_SGR_burst}. 
The total observed DM of an FRB comprises of contributions from the Milky Way's Interstellar Medium (hereafter ISM) (DM$_{\rm ISM}$), the Milky Way's Halo (DM$_{\rm HALO}$), the diffuse Intergalactic Medium (hereafter IGM) (DM$_{\rm IGM}$), the host galaxy's halo and ISM (DM$_{\rm HG}$) and the circumburst environment (DM$_{\rm SOURCE}$). Broadly, these can be grouped into the DM contribution of the Milky Way (DM$_{\rm MW}$), and the contribution of all extra-galactic components (DM$_{\rm EG})$, where:
\begin{equation}
    \mathrm{DM}_{\rm MW} = \mathrm{DM}_{\rm ISM} + \mathrm{DM}_{\rm HALO}, 
\end{equation}
and
\begin{equation}
    \mathrm{DM}_{\rm EG} = \mathrm{DM}_{\rm IGM} + \frac{\mathrm{DM}_{\rm HG} + \mathrm{DM}_{\rm SOURCE}}{(1 + z)}.
\end{equation}

In addition to the dispersion, a cold ionised plasma also causes scattering of the radio waves and results in multi-path propagation. This scattering manifests itself as (i) temporal broadening due to the multi-path propagation, (ii) scintillation bands due to the interference of scattered images, and (iii) angular broadening of the apparent source size. 
For impulsive signals such as FRBs, the effects of temporal broadening and scintillation can often be readily measured, while for extragalactic continuum radio sources the angular broadening is the most prominent effect of scattering.


Majority of the FRBs detected so far have shown exponential scattering tails and/or what appear to be scintillation bands in their spectra (\citealt{Cordes&Chatterjee2019_FRB_review, First_CHIME_catalog_2021, Macquart2019_spec_idx}). For almost all FRBs, the observed scattering is much larger than the expected scattering from these models of the Milky Way's ISM. The scattering properties of FRBs are thus a very useful probe of the turbulence in the plasma beyond the Milky Way, and lying along the line-of-sight to the FRBs.

Multiple theoretical studies have discussed the potential of using FRBs as probes of extragalactic turbulence \citep{Macquart&Koay2013, Cordes2016_FRB_scattering, Prochaska_Neeleman_2018_DLAs, Vedantham2019_scattering_CGM, Zhu_Feng_2018_scattering_hydrosims}. The dominant source of observed scattering in FRBs is expected to be external to the Milky Way, but the relative contributions from the host galaxies, IGM and galactic halos present along the line of sight remains unresolved \citep{Cordes&Chatterjee2019_FRB_review}. The largest reservoir of ionised plasma encountered by FRBs lies in the IGM, as is evidenced by the recent discovery of a DM$-$redshift($z$) relation in FRBs \citep{Macquart2020_IGM_Baryons_DM_z_relation}. If the ionised plasma in the IGM is also the dominant contributor to the observed scattering, the measured scattering properties of FRBs will predominantly probe the IGM, and potentially yield a relationship between DM and the scattering timescale, $\tau$. 
\cite{Ravi2019_observed_prop} searched for such a relation in FRBs using the sample of bursts detected with Parkes radio telescope and found evidence in support with a low to moderate level of significance. On the contrary, similar efforts by \citealt{Hao_qiu_2020_ASKAP_FRB_scattering_props} and \citealt{Cordes2016_FRB_scattering} have ruled out such a relation. If it exists, a DM-$\tau-$ relation would be critical in establishing whether the dominant source of scattering and dispersion in FRBs is the same \citep{Macquart&Koay2013}. 

Alternatively, \cite{Vedantham2019_scattering_CGM} have suggested that the Circumgalactic Medium (CGM) of intervening galaxies could explain scattering timescales of the order of milliseconds (at 1 GHz), and therefore measurements of temporal broadening in localized fast radio bursts can be used to constrain the properties of the cool ionized gas clumps in the CGM of intervening galaxies. However, \cite{Prochaska2019_FRB181112} measured a scattering timescale (which was later refined by \cite{Cho2020_pfbinverted_181112}) of $\lesssim 20~\mu$s at 1.3 GHz in FRB20181112A (which had been found to intersect the halo of a foreground galaxy) allowing them to place constraints on the density and turbulence of the ionised plasma in the halo of the foreground galaxy. Similarly, scattering timescales reported for repeating FRB sources FRB20180916B and FRB20121102A were used by \cite{Ocker_and_Cordes_2021_Halo_scattering_limits} to limit the scattering contribution from the Milky Way Halo to an FRB's scattering budget to values less than 12 $\mu$s.
More recently, \cite{Chawla2021_scattering_CHIME_FRBs} performed a population synthesis analysis using the properties of FRBs reported in the first CHIME/FRB catalog \citep{First_CHIME_catalog_2021} and found that a model where scattering originates in the turbulent medium local to the FRB, combined with the circumgalactic medium of intervening galaxies is consistent with the observed properties of their FRB sample.

Therefore, the observed signatures of scattering imprinted upon the FRB profiles are salient features which enable the use of FRBs as a cosmological probe. A caveat however, is that the incoherent dedispersion search technique implemented by most FRB search pipelines precludes the accurate measurement of the scattering widths of FRBs. 
Access to the raw voltage data, which typically requires a real-time detection system, enables the use of coherent techniques of removing the dispersion and a near-perfect reconstruction of the intrinsic burst profile unaffected by instrumental smearing.

In this paper, we report the detection of FRB20191107B with the UTMOST radio telescope, and describe its remarkably narrow intrinsic width and scattering timescale revealed after applying coherent dedispersion to the raw voltage data captured for the FRB. Section \ref{sec: FRB191107B props} describes the methodology we use to model the burst properties. In Section \ref{sec:rates of narrow FRBs} we discuss the rates of intrinsically narrow FRBs and the efficiency of current leading surveys in probing the population of such FRBs. In Section \ref{sec: IGM scattering props} we use measured scattering properties of FRB20191107B to put constraints on the strength of turbulence in the IGM and search for a DM$-\tau$ relation in FRBs. In Section \ref{sec: Origin of scattering} we discuss the potential dominant regions for the origin of the observed scattering and identify the local environment of the source as the most likely candidate. We summarise and make our conclusions in Section \ref{sec: Conclusions}.

\section{Detection of FRB20191107B}
\label{sec: FRB191107B props}

\begin{figure}
    \centering
    \includegraphics[width=0.45\textwidth]{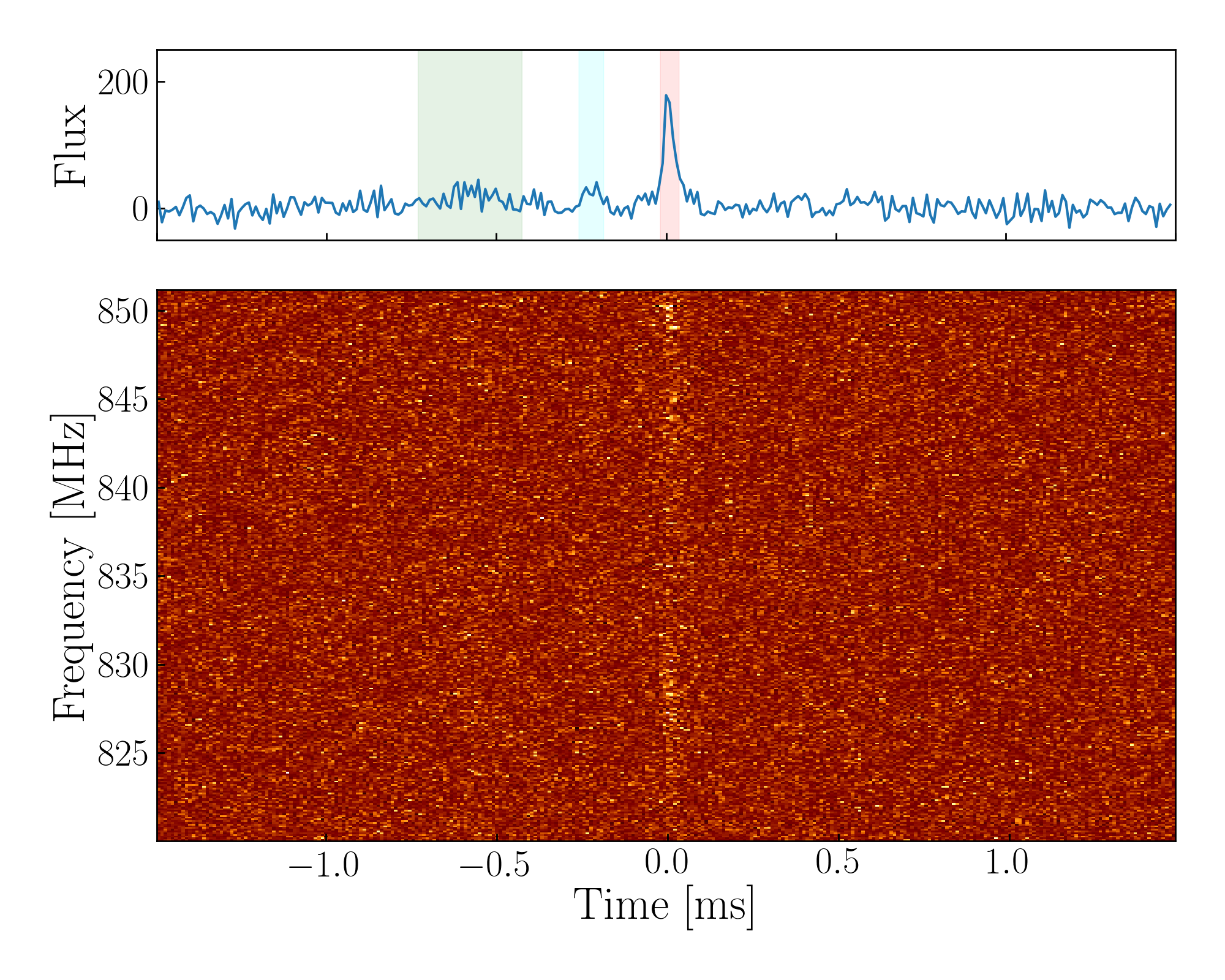}
    \caption{The dynamic spectrum (bottom) and frequency averaged intensity time series (top) of FRB20191107B, after correcting for the dispersive delay using coherent dedispersion. The capability of UTMOST to capture raw voltages at the native resolution of the instrument has revealed the intrinsically narrow width and the weaker components of the burst. The $x$-axis spans $\sim 3$ milliseconds of data, and each time sample is 10.24 $\mu$s wide. Three components have been identified which have been highlighted in green, cyan and red intervals in the top panel. While they look very weak to the eye, their detection is statistically significant (see Section \ref{subsec:modelling_191107}) and becomes prominent when the data are averaged.}
    \label{fig:FRB191107 }
\end{figure}

UTMOST is a 1.6 km long cross radio-interferometer located in New South Wales, Australia \citep{utmost}. Operating at a centre frequency of 835 MHz, it has been running multiple FRB search surveys over the past 5 years (\citealt{Caleb_3frbs, Farah2019}; Gupta et al. in prep.). UTMOST uses a machine learning based real-time detection and classification pipeline \citep{Farah2019} and has discovered 18 FRBs so far. 

FRB20191107B was detected in real-time with our FRB pipeline at UTMOST, and raw voltage data sampled at the Nyquist rate of the receiver instrument were captured for $\sim$800 ms around the event.

The FRB was initially detected at a DM of 715.7 \dm\ and with a Signal-to-Noise (S/N) ratio of 9.9. The observed width was 1.3 ms but due to the event's high DM, this is dominated by the smearing of the pulse due to intra-channel dispersion smearing at the 0.097 MHz frequency resolution of the instrument. 
The captured voltages not only provide access to the full-time-resolution data (10.24 $\mu$s), but also preserve the full phase information, allowing for coherent dedispersion of the burst. This revealed the FRB to be a bright and narrow pulse with even narrower components. 
We used the \texttt{pdmp} tool from the \texttt{PSRCHIVE}\footnote{\url{http://psrchive.sourceforge.net}} software to optimise the burst's DM and S/N using the high time resolution data and measure a S/N of 23 at a DM of 714.9 \dm\ and a box-car width of only 61 \us.
We reprocessed the voltage data and coherently dedispersed the burst at the optimised DM reported by \texttt{pdmp}, which revealed three individual narrow components with a hint of a scattering tail (Fig \ref{fig:FRB191107 }). We model and use the resulting profile for the analysis that follows. The discovery of the FRB was promptly reported as an Astronomer's Telegram in \cite{GuptaAtel2019c_FRB191107} to allow for rapid multi-wavelength follow-up.

Following the methodology used in \cite{Farah2019}, we use the radiometer equation to estimate the apparent fluence of the burst at 6.7 Jy ms. Due to the co-linear arrangement of the individual elements on the East-West arm of the telescope there is a large uncertainty in the localisation of the FRB in the North South direction ($\sim 2 \deg$), and the location of the burst within the primary beam of UTMOST cannot be constrained, such that the fluence is a lower limit. The localisation arc of the burst is an elongated ellipse and can be described using the following equation: 
\begin{multline}
RA = 8.032153 - 2.313314\times10^{-4} \times (DEC + 13.837823) \\ 
  + 1.009132\times 10^{-5} \times (DEC + 13.837823)^{2}
\end{multline}

where $RA$ is in hours, $DEC$ is in deg, and is valid in the range $DEC$ = [-17.1, -10.6]. The best fit coordinates of the FRB are: $RA$ = 08:01:57.08, $DEC$ = -13:44:15.5.

\subsection{Modelling the burst properties}
\label{subsec:modelling_191107}

\begin{figure*}
    \centering
    \includegraphics[width=0.99\textwidth]{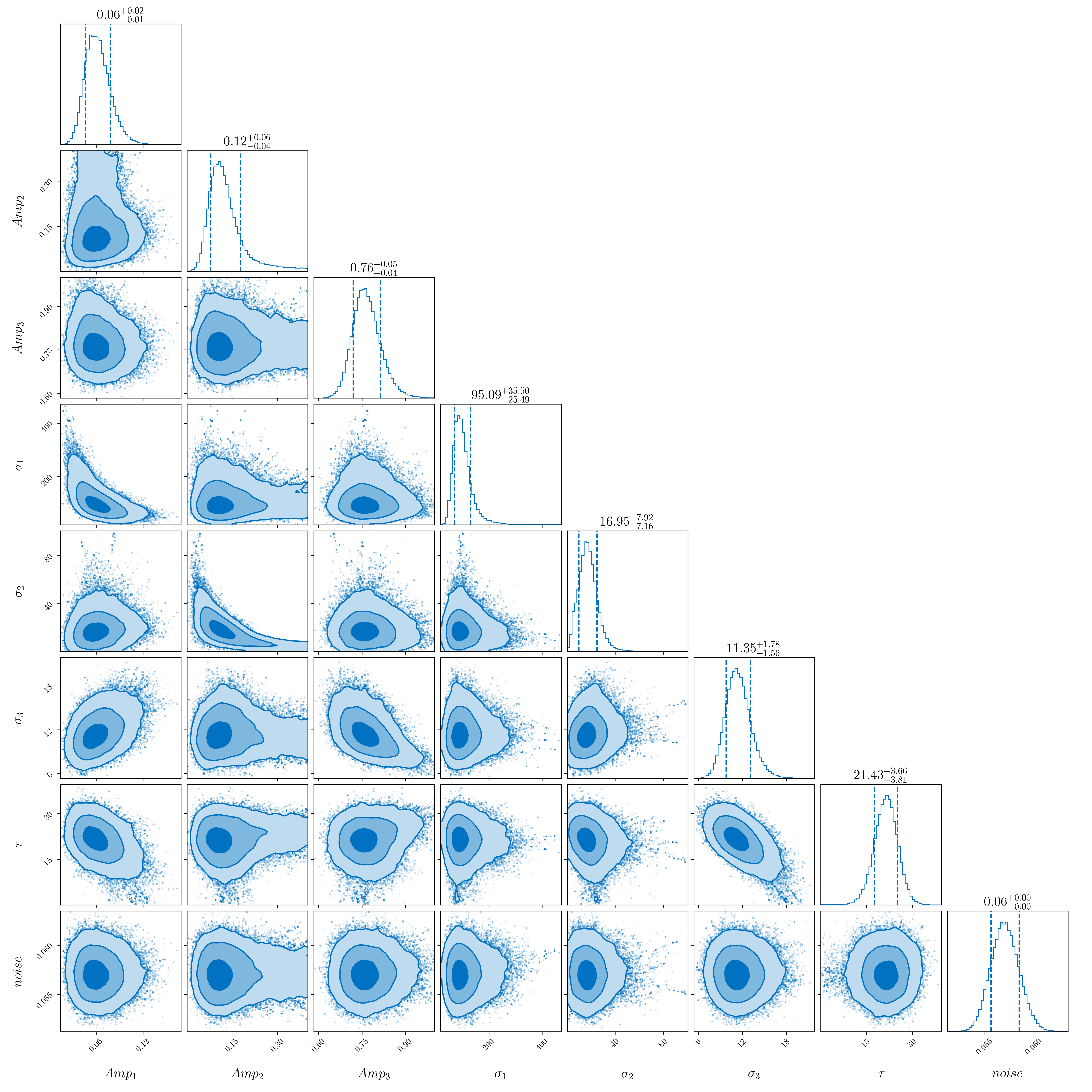}
    \caption{The joint posterior distribution of model parameters in Eqn \ref{fittingeqn}. The dashed lines represent 16 and 84 percentiles in the marginalised 1-D histograms, and the best fit values of the parameters are listed on the top of each histogram.}
    \label{fig:posteriors}
\end{figure*}

\begin{figure}
    \centering
    \includegraphics[width=0.45\textwidth]{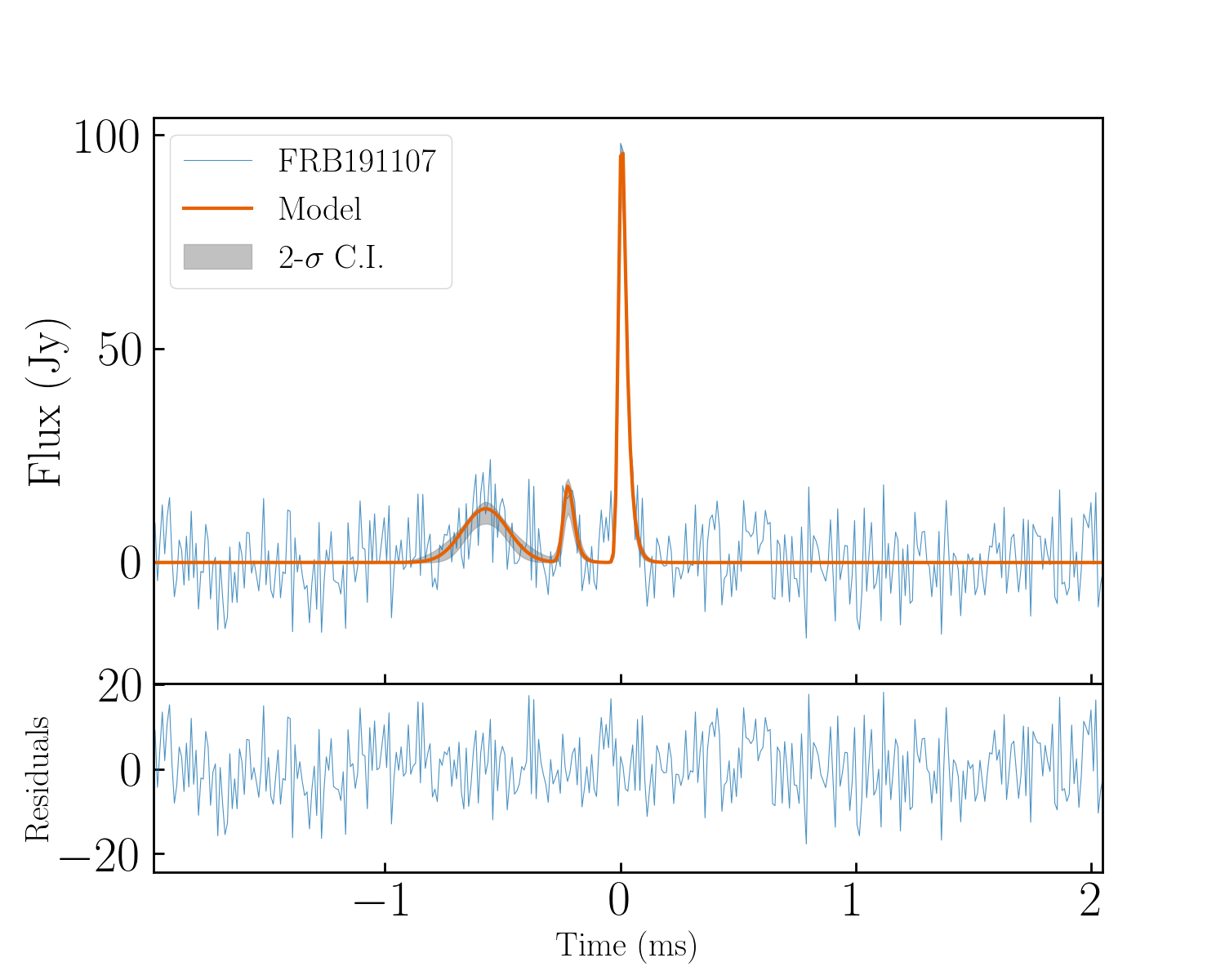}
    \caption{The pulse profile of FRB20191107B showing its three individual components. The best fit model is plotted in orange along with 2-$\sigma$ contours shaded in black. The bottom panel shows the residuals after subtracting model from the data.}
    \label{fig:model_fit_profile }
\end{figure}

The burst profile comprises of three sub-bursts with a trailing, bright narrow component and a hint of an exponential tail as is typical for a radio signal propagating through turbulent media and undergoing multi-path propagation. We therefore model the profile as three Gaussian pulses convolved with a one sided exponential of the form:
\begin{equation}\label{fittingeqn}
    S=\sum_i^{n=3} \left\{ \mathrm{A}_i \times \exp \left[\frac{-\left(t-t_i\right)^2}{2 \sigma_i^2}\right] \right\} *\left\{\exp \left[-\frac{t}{\tau}\right]\right\},
\end{equation}

 \noindent where * denotes convolution. Here, $\tau$ is the scattering timescale, $\sigma_i$ denotes the width of the Gaussians, $\mathrm{A}_i$ are the amplitudes, and  $t_{i}$ are the centres of the Gaussians. 
 Parameter estimation was performed using the \texttt{BILBY} package \citep{Bilby}, making use of the in-built Markov Chain Monte Carlo (MCMC) sampler \citep{emcee_v3}.
 
 For parameter estimation, we start with uniform priors on all fitted parameters, and use a Gaussian log$-$likelihood function ($L$) defined as:
 \begin{equation}
     L = -0.5 \times \left[\sum_{j} \left( \frac{(D(t_{j}) - S(t_{j}))^{2}}{n^{2}} \right)+ \log(2 \pi n^{2})  \right]
 \end{equation}
 
 \noindent where $D$ is the frequency-averaged, time-series data, $S$ is the model, and $n$ is an estimate of the noise per time sample.
 
 Following a burn-in stage, the joint posterior density of all parameters was estimated with 9,000 samples. The resulting distribution of the posteriors is shown in Fig \ref{fig:posteriors}. 
 We find that the data are well modelled with the single Gaussian components convolved with an exponential. To test the significance of detection of scattering and detection of the first two weaker components, we compute the Bayes Factor ($\mathcal{B}$) for our model and compare it with the models where we exclude the scattering parameter ($\tau$) and/or the parameters corresponding to the first two Gaussian components ($\sigma_{0,1}, A_{0,1}, t_{0,1})$. We find that our model with the three components and the scattering term included is strongly favoured by the data with $\log \mathcal{B} > 5$ \citep{jeffreys1998theory_bayes_factor_citation, Trotta_2008_Bayes_factor_Jeffreys_scale}. We summarise the Bayes Factor values of our models against a model with a single component with no scattering in Table \ref{tab: model evidences}.
 
\begin{table}

\centering
\begin{tabular}[c]{|ccc|}
\hline
\hline
Number of components & With scattering tail & Without scattering tail \\
\hline
Bursts 1, 2 and 3 & 32.47 & 27.38 \\
Bursts 2 and 3 & 14.35 & 17.74 \\
Burst 3 only & 7.34 & 0 \\
\hline
\end{tabular}

\caption{This table lists the log Bayes Factor ($\log \mathcal{B}$) values of different models fit to the FRB profile as compared to a model with only one burst component with no scattering. This table shows that our model with three individual components along with an exponential scattering tail provides the best fit to the data, and that the existence of the two weaker components is statistically significant ($\log \mathcal{B} > 5).$}
\label{tab: model evidences}
\end{table}

 Using the best fit model we measure a scattering time ($\tau$) of $21.4^{+4}_{-3}$ \us\ and widths($\sigma_i$) of the three Gaussians as $95.0 ^{+35}_{-25}$, $17.0^{+8}_{-7}$ and $11.3^{+3}_{-3}$ \us, where the reported uncertainties are the 68\% confidence interval. These values indicate that this FRB consists of one of the narrowest components in an FRB detected with UTMOST until now. However, it is worth mentioning that repeat bursts from FRB20180916B and FRB20200120E have previously been observed to show components with narrower widths \cite{Nimmo2021_microstructure_180916, Nimmo2021_nanosecond_FRB200120E}. The three components are found to be separated in time with a gap of 360 \us\ between the first and the second components, and of 230 \us\ between the second and the third components (see Fig \ref{fig:model_fit_profile }), suggesting that the emission regions associated with each component would only be a few kilometers in size.
 
 In the next section, we discuss the significance of the narrow width of the brightest component of the FRB and the sensitivity of FRB surveys to such FRBs. The remarkably low scattering time despite a relatively large observed DM provides the opportunity to place limits on the strength on turbulence in the Intergalactic Medium (IGM) along the line-of-sight to this source, which we explain in Section \ref{sec: IGM scattering props}.

\section{Rates of narrow FRBs}
\label{sec:rates of narrow FRBs}
\begin{figure}
    \centering
    \includegraphics[width=0.47\textwidth]{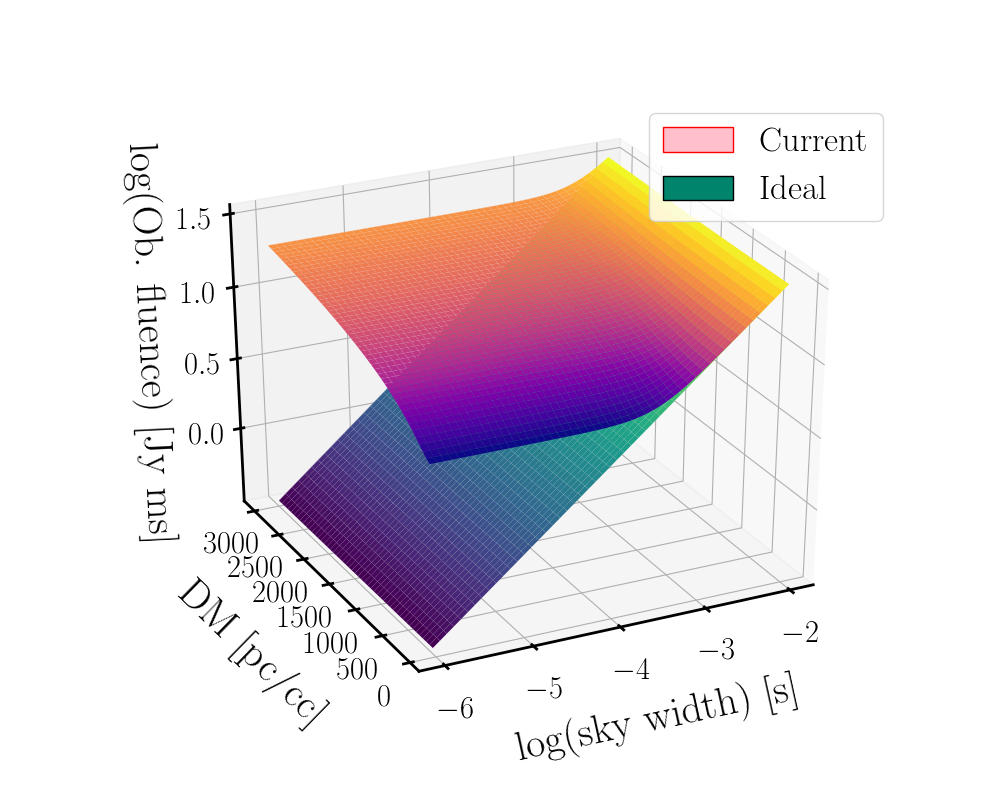}
    \caption{Detection fluence threshold of UTMOST as a function of the sky-width and DM of FRBs. The plane in orange-yellow shows the detection threshold with the current time and frequency resolution of the instrument, and the green-blue plane shows the detection threshold if UTMOST had infinitely high resolution, i.e. for an ideal instrument. }
    \label{fig:Fluence_threshold_UTMOST}
\end{figure}

\begin{figure}
    \centering
    \includegraphics[width=0.45\textwidth]{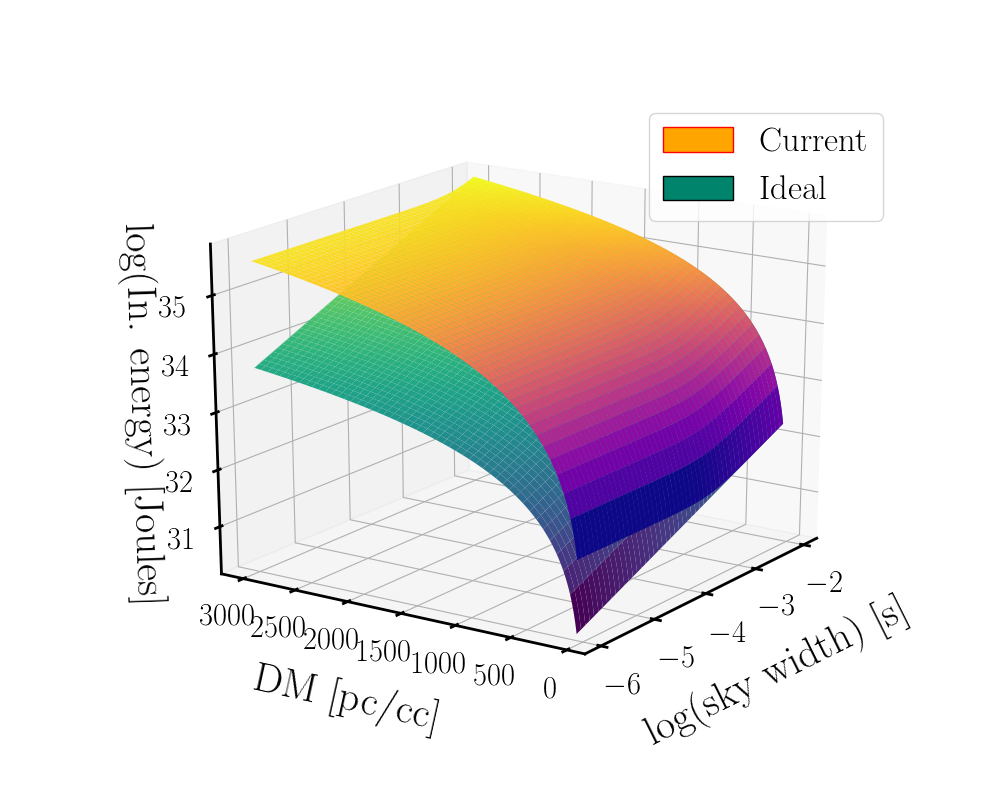}
    \caption{Planes of intrinsic isotropic energy threshold as a function of the intrinsic width and DM of FRBs, for the current survey (orange-yellow) as well as an ideal survey (blue-green) with UTMOST. Our current survey with UTMOST would not be sensitive to FRBs originating in the region between the two planes, which would be detected by an ideal survey with infinite time and frequency resolution. Strictly speaking, our thresholds are upper-limits as our redshift estimates from the DM values are upper-limits, which are used in computing the intrinsic isotropic emission energy in Equation \ref{eqn: energy thresh}. However, correcting for source redshift would equally affect both the `Ideal' and the `Current' survey equally, keeping intact the gap between the two planes.}
    \label{fig:Energy_threshold_UTMOST}
\end{figure}

\begin{figure}
    \centering
    \includegraphics[width=0.37\textwidth]{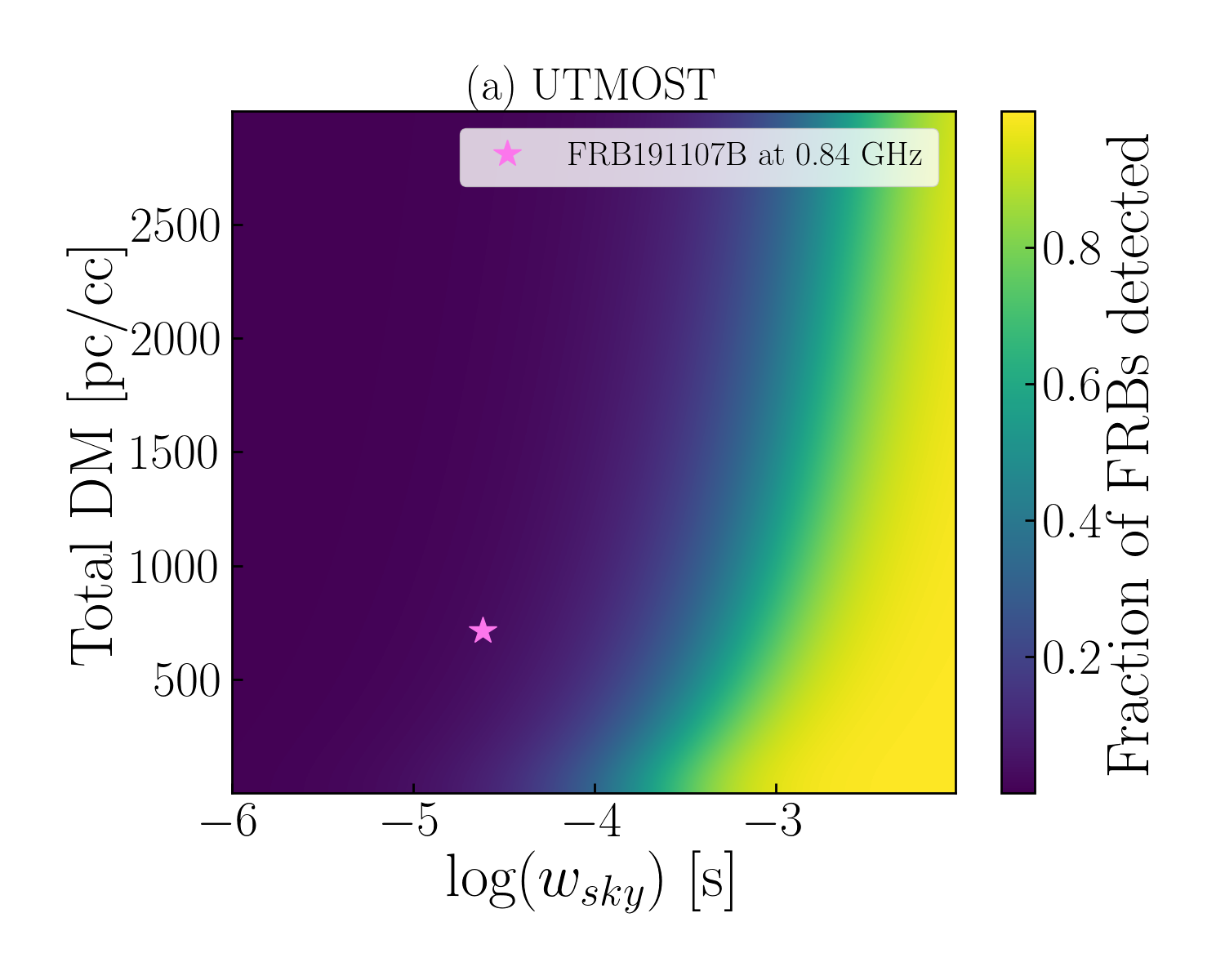}
    \includegraphics[width=0.37\textwidth]{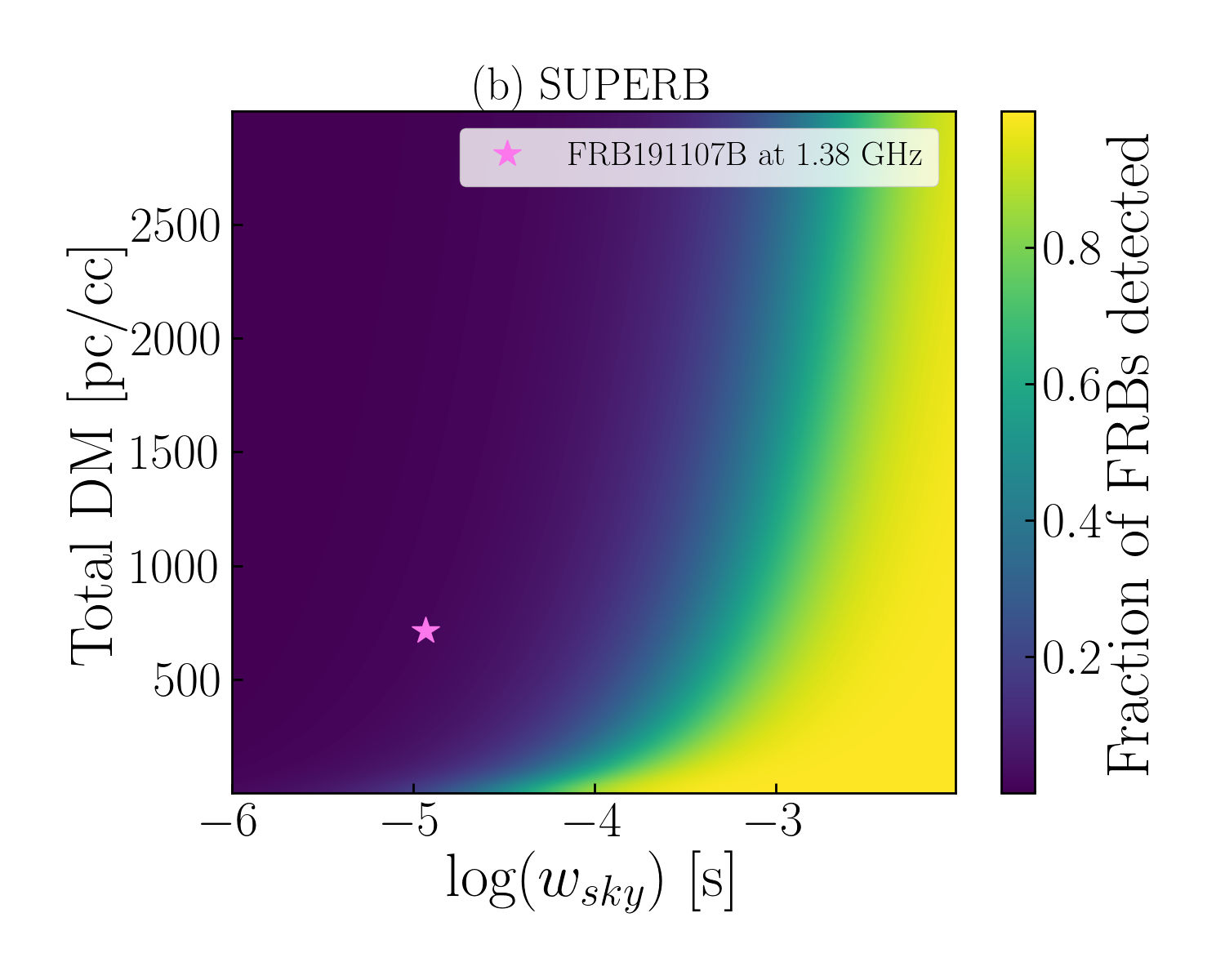}

    \includegraphics[width=0.37\textwidth]{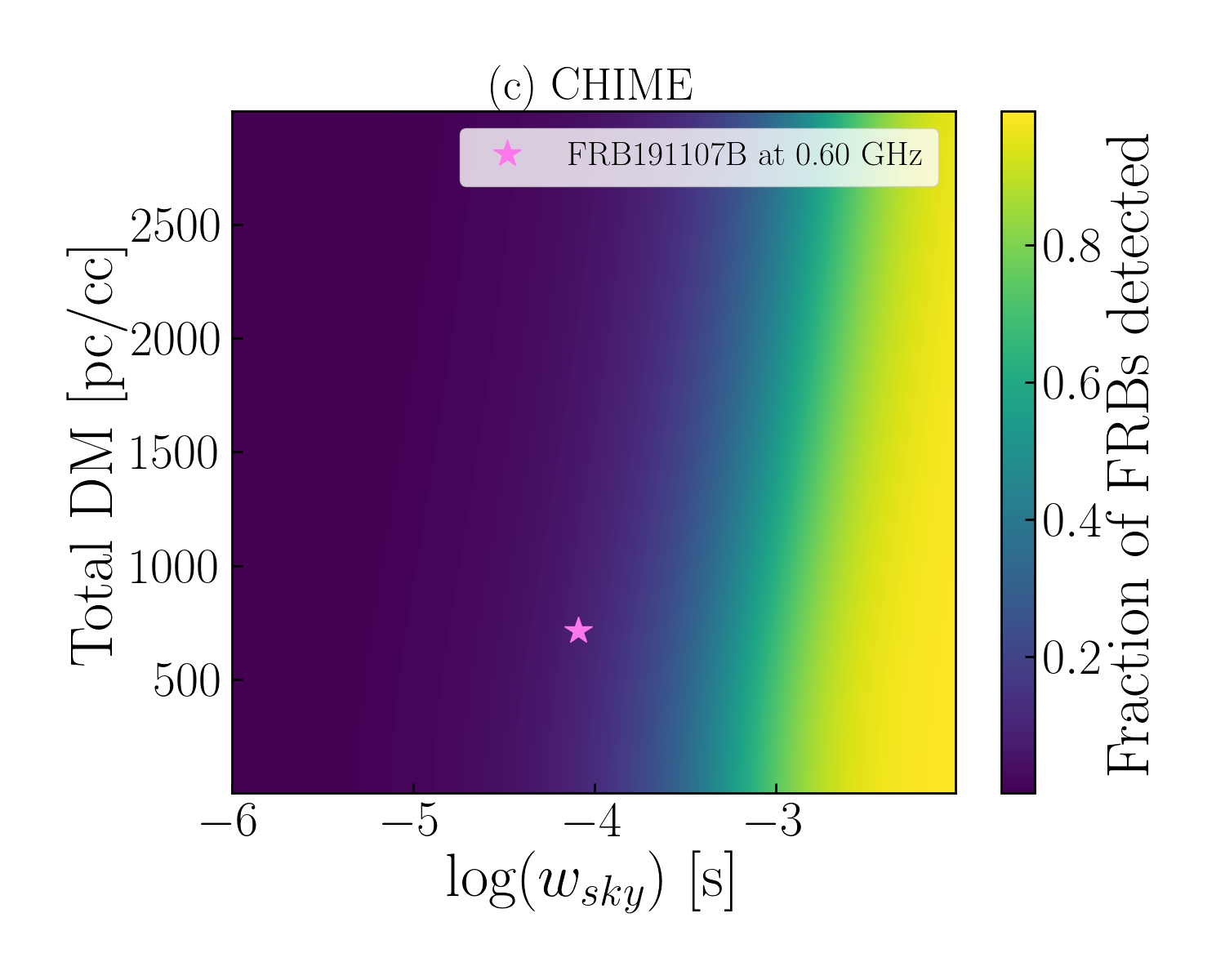}

    \includegraphics[width=0.37\textwidth]{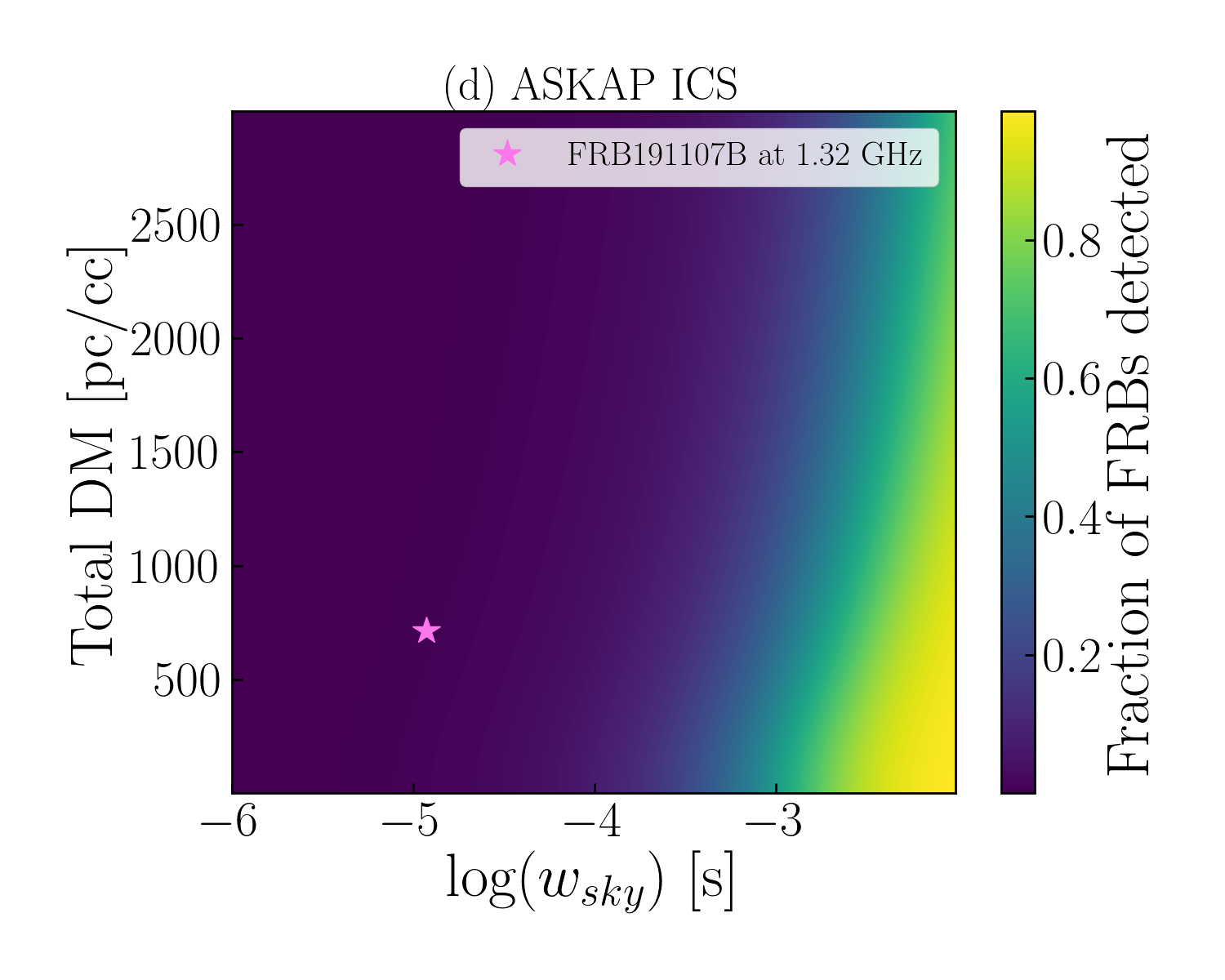}

    \caption{The distribution of the fraction of the FRB population that would be detected by a survey as a function of the intrinsic FRB sky-width and DM. We have assumed that the FRB population at a given redshift has a power-law intrinsic energy distribution (see Section \ref{sec:rates of narrow FRBs} for details). Different panels show this distribution for 4 prominent FRB surveys. The pink star marks the sky-width ($w_{sky}$) and DM of FRB20191107B scaled to the central frequency of the corresponding survey in each panel. It is evident that all surveys are probing only a small fraction of the population of FRBs like FRB20191107B.}

    \label{fig:Det_fraction_surveys}
\end{figure}

\begin{figure}
    \centering
    \includegraphics[width=0.45\textwidth]{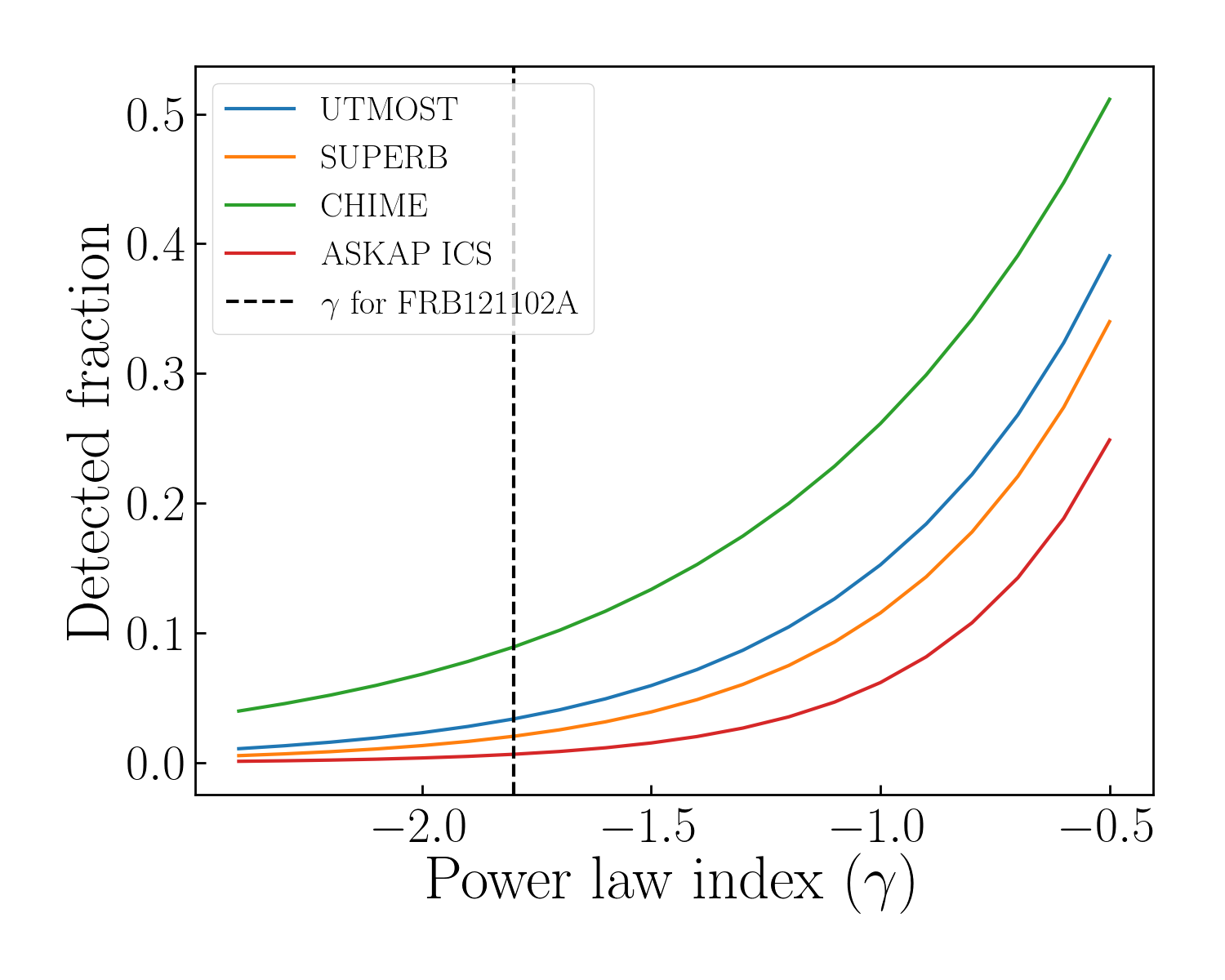}

\caption{The detection fraction of FRBs at FRB20191107B's sky-width and DM, as a function of the assumed power-law index $(\gamma)$ of the intrinsic energy distribution, for the UTMOST, SUPERB, CHIME and ASKAP (for ``incoherent sum'' mode) FRB surveys. The vertical dashed line indicates $\gamma = -1.8$, the power-law index of the intrinsic energy distribution of repeat bursts from FRB20121102A measured by \citep{Gourdji2019_121102_arecibo}, which is the value we adopt to make the maps presented in Fig \ref{fig:Det_fraction_surveys}.}

\label{fig:Det_fraction_UTMOST}

\end{figure}
Despite being intrinsically bright at microsecond resolution, FRB20191107B was detected with a S/N of 9.9 during the real-time search, just above our detection threshold of 9. 
The decrease in S/N of the intrinsically narrow burst when observed at coarser time resolution is expected, as, for a constant fluence, the S/N ratio of a burst is inversely proportional to the square-root of its observed width ($w_\mathrm{obs}$). Scattering from the ionised turbulent plasma along the propagation path of the burst can cause the intrinsic width of the burst ($w_i$) to be broadened with an exponential decay timescale ($\tau$). The total width of the burst at the time it reaches the telescope ($w_\mathrm{sky}$) can be calculated as
\begin{equation}
\label{eqn:tsky}
    w_\mathrm{sky} = \sqrt{w_i^2 + \tau^2}. 
\end{equation}

The observed width of the burst may additionally be smeared by the resolution of the recording instrument. 
Additionally, the sampling time of the data searched for bursts in real-time can be much higher than the sky-width of FRBs. The final observed width of a burst with an instrument with a sampling time $w_s$ can be estimated using the following relation:
\begin{equation}
\label{eqn:tobs}
    w_{\rm obs} = \sqrt{w_{\rm sky}^2 + w_s^2 + w_{\rm DM}^2},
\end{equation}

\noindent where $w_{\rm DM}$ is the width due to the intrachannel DM smearing based on the frequency resolution of the instrument, and is calculated as (see \citealt{Pulsar_Handbook_2004_Lorimer_and_Kramer}):
\begin{equation}
\label{eqn:tdm}
    w_{\rm DM} = 8.3 \times 10^{-6} \times \Delta \nu \times {\rm DM} \times \nu_c^{-3} ~s,
\end{equation}
\noindent and $\Delta \nu$ is the width of each frequency channel in MHz and ${\nu}_c$ is the central frequency of the instrument in GHz.

For a given fluence, the flux density of the source increases with decreasing width. This leads to a boost in the measured S/N of bursts and therefore decreases the sensitivity threshold of an instrument towards intrinsically narrow sky-width events.
However, this boost in S/N only happens as long as the width of the burst is larger than the sampling time of the recording instrument or the DM smearing width due to its finite channel bandwidth, whichever is greater. Therefore, bursts with even narrower widths do not benefit from their increased flux.
As a result, the fluence detection threshold of a survey does not decrease with the width and weaker but intrinsically narrow events which could have been detected by the instrument remain undetected. 

Previous work by \cite{Connor2019_interpreting_the_distribution_of_frb_observables} on the interpretation of the width and DM distributions of FRBs detected with different telescopes, has highlighted the possibility that a population of narrow width FRBs may exist which remains relatively unexplored by the current surveys. In order to investigate the fraction of FRB population UTMOST would miss due to this limitation in the sensitivity of our survey, we model a 3-dimensional plane of our survey's detection threshold as a function of the sky-width, DM and the observed fluence of FRBs. Using equations \ref{eqn:tsky}, \ref{eqn:tobs}, \ref{eqn:tdm}, and the reported value of System Equivalent Flux Density (SEFD) we calculate the detection fluence threshold of UTMOST over a 2-dimensional grid of sky-width and DM of FRBs. This plane of detection fluence threshold is shown in Fig \ref{fig:Fluence_threshold_UTMOST} (labelled as `Actual'). We also show the plane of the detection fluence threshold of a hypothetical ideal survey with UTMOST which has infinitely narrow time and frequency resolution such that it does not suffer from any of the instrumental smearing effects on the detected width of the bursts (labelled as `Ideal'). The gap between the two detection threshold planes (actual and ideal) represents the region of incompleteness where our survey would not be detecting any FRBs even though it lies above the theoretical sensitivity limit of the telescope. It is worth mentioning here that the `Actual' threshold is valid only for an ideal detection pipeline, and the search pipeline could have other selection biases against narrow bursts, which are not characterised by Equations \ref{eqn:tsky}--\ref{eqn:tdm}. We do not perform absolute calibration of selection effects through injections, so it is possible that additional narrow bursts are being missed by the survey.

As is now well demonstrated, the DM of FRBs correlate strongly with redshift, \citep{Macquart2020_IGM_Baryons_DM_z_relation}, allowing us to convert the observed specific fluence of bursts into the intrinsic isotropic emitted energy at source using the relation \citep{BingZhang2018_detectibility_of_high_z_FRBs}:
\begin{equation}
    \label{eqn: energy thresh}    E_{\rm int, iso} = \frac{F_{\rm obs}}{(1+z)} \times 4 \pi D_L^2 \times \nu_c 
\end{equation}
\noindent where $F_{\rm obs}$ is the observed specific fluence, $\nu_c$ is the central frequency of detection, and $D_L$ is the luminosity distance to the source. To avoid unnecessary complexity and edge cases we take the IGM's DM contribution as approximately equal to the total DM of an FRB, and use a simple linear relation between DM and source redshift ($z$) ($DM = 1000 \times z$ \dm) to approximate the redshift of the source (upper-limit), from which we estimate the luminosity distance assuming a standard $\Lambda$-CDM cosmology. This allows us to transform the fluence detection threshold planes (actual and ideal) into the detection threshold planes for the intrinsic, isotropically emitted energies of bursts, as a function of the DM and sky-widths. These planes are shown in Fig \ref{fig:Energy_threshold_UTMOST} for the UTMOST's FRB survey.

If we assume the cumulative intrinsic energy distribution of FRBs at a given redshift follows a power-law function ($N(>E_{\rm int}) \propto E_{\rm int}^\gamma$) with an index $\gamma$, we can then calculate the completeness fraction of FRBs detected with a given survey as the ratio of the number of FRBs above the plane of actual detection threshold to the number of FRBs above the plane of ideal detection threshold. This completeness fraction quantifies the efficiency with which a given survey probes the narrow-width FRB population which lies above its quoted fluence/energy detection threshold. Here we have assumed that the low-energy cutoff of the power-law distribution of FRB intrinsic energy lies below the detection thresholds of all surveys.

We compute this fraction for a few prominent FRB surveys like SUPERB, CHIME/FRB, UTMOST and ASKAP (InCoherent Sum Survey) as functions of sky-width and DM. We assume an intrinsic energy power-law index of $-1.8$ based on the measured value for FRB20121102A by \cite{Gourdji2019_121102_arecibo}. These maps of detected fraction of FRBs are shown in Fig \ref{fig:Det_fraction_surveys}. It is worth highlighting here that this bias against narrow width bursts shown in Fig \ref{fig:Det_fraction_surveys} is due to the frequency and time resolution of the detecting instrument and the selection effects due to inefficiencies in the search pipeline (such as 0-DM RFI excision, poorly trained machine learning based classifier) would be in addition to those presented here. Therefore, our plots do not show a decrease in burst recovery fraction at higher widths as presented by \citealt{Gupta_et_al_2021_mock_frb_injection_utmost} and \citealt{First_CHIME_catalog_2021}.

It is evident from these maps that all surveys would detect only a small fraction of the population of FRBs with narrow sky-widths. For reference, we also plot the measured value of the sky-width of FRB20191107B, after scaling the scattering time to the center frequency of each telescope. 

The measured detection fraction is strong function of the intrinsic source energy distribution, and we show this dependence by plotting the detection fraction for a given survey at FRB20191107B's sky-width and DM as a function of the assumed power-law index $\gamma$ for the source energy distribution in Fig \ref{fig:Det_fraction_UTMOST}.

In summary, we find that most ongoing surveys could be missing $>$60\% of the population of FRBs at the observed DM and sky-width of FRB20191107B. The fact that UTMOST detected one FRB with such narrow sky-width and a relatively large DM, suggests that there might exist a significant population of FRBs with narrow intrinsic widths and small scattering times, which remains largely unexplored with the current surveys.

\section{FRB20191107B and properties of the IGM}
\label{sec: IGM scattering props}

FRB20191107B shows a scattering timescale of only 21 $\mu$s despite having a relatively large DM of $\sim 715$ \dm, offering an interesting source from which to constrain the scattering properties of the IGM. The NE2001 model \citep{NE2001} estimates a contribution from the Milky Way's Interstellar Medium (ISM) of 127 \dm. This model does not include the effects of the Milky Way halo, and increasingly, this correction to the FRB DM has been adopted by recent works. For example, \cite{Prochaska2019_haloes} have suggested that the halo of Milky Way contributes between 50$-$80\dm\ to the DM, while \cite{Bhardwaj2021_M81_frb_repeater_frb20200120E} have suggested an upper limit of $<$53\ \dm\ from the Milky Way halo (based on an FRB found in the very nearby M81 galaxy). 
Here we adopt 50 \dm\ as the DM contribution from the MW halo in all directions on the sky. 
Subtracting 50 \dm\ for the Milky Way's halo contribution, we are left with a DM excess (DM$_{\rm EG}$) of 537 \dm\ for FRB20191107B which we attribute to the cumulative contributions from the IGM, intervening halos and the host galaxy of the FRB. 

\cite{Macquart2020_IGM_Baryons_DM_z_relation} have shown that the IGM is the dominant source of DM for FRBs coming from high redshift galaxies, and that DM can be used as a proxy for distance to the host galaxy of FRBs. 
The DM contribution from the IGM and intervening halos can be estimated approximately from the source redshift by \citep{Inoue, Ioka2003, Macquart2020_IGM_Baryons_DM_z_relation}:
\begin{equation}
    \label{eqn:DM-z relation}
    \mathrm{DM}_\mathrm{IGM} \approx 1000 \times z ~~~{\rm pc\,cm}^{-3}
\end{equation}

We adopt this relation to estimate an upper limit on the redshift of 0.53 for the host galaxy. 

\subsection{Scattering in the IGM}
\label{subsec:Scattering_IGM}

Impulsive radio signals originating from cosmological distances provide a unique tool to probe the turbulent properties of the IGM in exquisite detail. FRBs carry on them the signature of the properties of the ionised plasma they have travelled through along their propagation path.

The scattering strength of an ionised medium is quantified by the Scattering Measure (SM), defined as (see \citealt{Cordes_and_Lazio_1991_Scintillation_theory}): 
\begin{equation}
    {\int_{L}^{L+\Delta L}C_N^2 ~dl}, 
\end{equation}
where $C_N^2$ is the amplitude of turbulence per unit length of the plasma extended between L and L$ + \Delta$L. 
It is usual to simplify by modeling the inhomogeneities associated with the plasma to be located in a single plane, known as the scattering screen. 
This thin-screen approximation is a valid assumption in the scenario when one turbulent region dominates the inhomogeneities along the propagation path.
Nevertheless, it is common to apply this assumption in case of extended turbulent medium as the effects of an extended medium can be treated as effects of an equivalent thin-screen with some modified parameters such as the effective screen distance and the strength of turbulence \citep{Lee_and_Jokipii_1975_scattering_theory}.

\cite{Macquart&Koay2013} provide a relation between the observed scattering timescale ($\tau$) and the SM of the intervening medium for applications related to FRBs. They incorporate the effect of the curvature of space-time due to expansion of the universe in the definition of scattering measure as:
\begin{equation}
{\rm SM_{eff}} = \int \frac{C_N^2(l)}{(1 + z)^2} dl. 
\end{equation}

The observed scattering timescale due to scattering of a radio pulse in the turbulent plasma and the ${\rm SM_{eff}}$ of the medium are related (when the diffractive scale is larger than the inner scale of turbulence in the scattering medium) as:
\begin{equation}
\label{eqn:tau-SM}
    \tau = 1.9 \times 10^{-4} ~ (1 + z_L)^{-1} {\rm  \bigg( \frac{\lambda_0}{1~m} \bigg) ^{22/5}    \bigg( \frac{D_{\rm eff}}{1~Gpc} \bigg) \bigg( \frac{SM_{\rm eff}}{10^{12} ~m^{-17/3}}\bigg)^{6/5} ~s}
\end{equation}

\noindent where $\lambda_0$ is the wavelength in the observer frame, $z_L$ is the redshift of the scattering-screen, and ${\rm D_{eff}}$ is the ratio of angular diameter distances $D_LD_{LS} / D_S$, where $D_L$, $D_{LS}$ and $D_S$ are the angular diameter distances to the scattering screen, screen to the source, and to the source respectively.

Temporal broadening due to scattering is modulated by the lever arm effect, which maximises the scattering mid-way between the source and observer ($D_L = D_S/2$) (for example in the IGM) as compared to a screen located near the observer or the source, as would be the case when the turbulence in the Milky Way ISM or in the host galaxy respectively.

\begin{figure}
    \centering
    \includegraphics[width=0.42\textwidth]{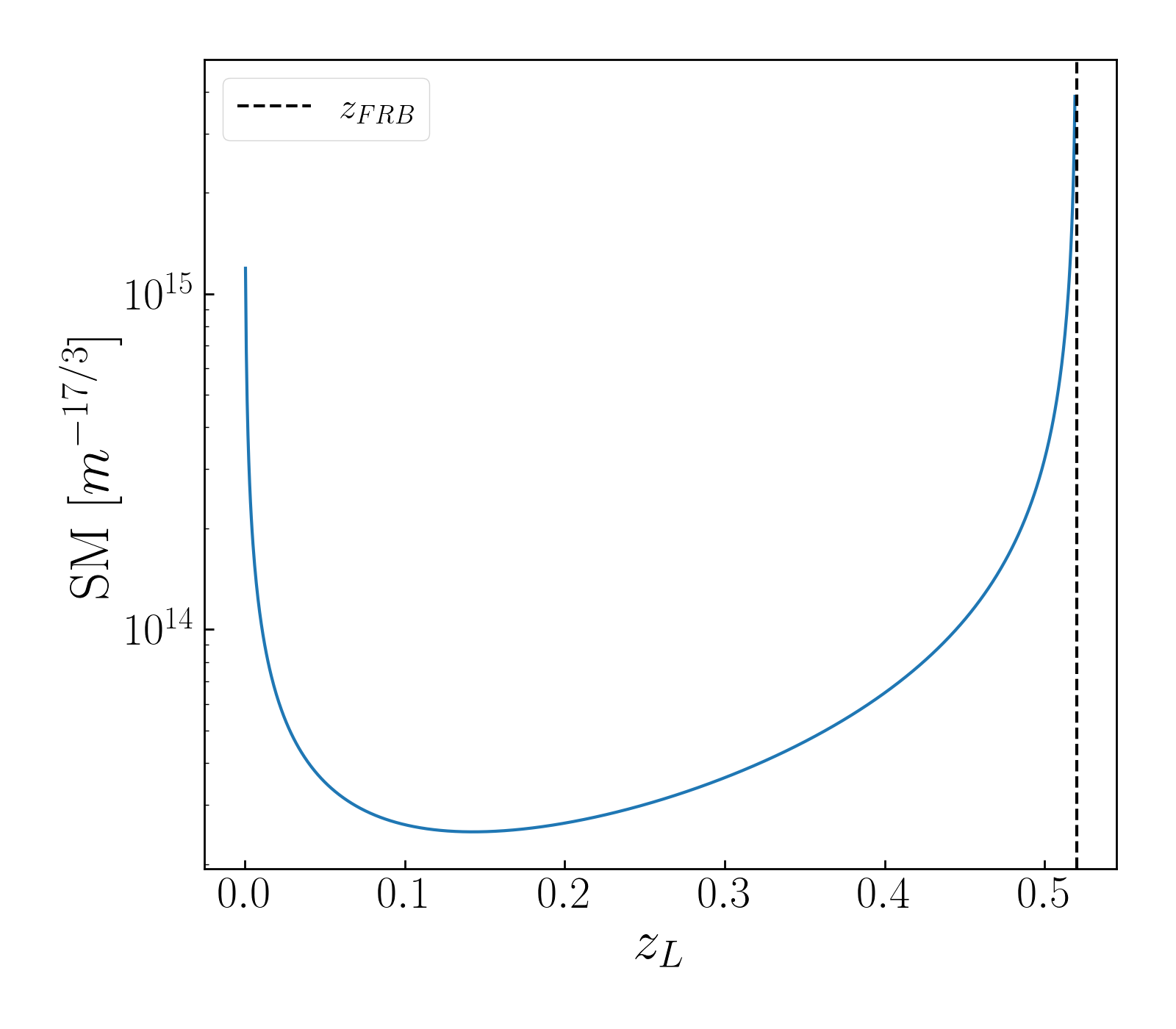}
    \caption{Upper limit on the scattering measure of the IGM as a function of the redshift of the screen ($z_L$). The geometric lever arm effect increases the scatter broadening of a pulse from plasma located equally far away from the source and the observer, reducing the requirement on the strength of turbulence of the IGM. The strongest limit of SM $< 2.6\times 10^{13} {\rm ~m^{-17/3}}$ (or $< 8.4\times 10^{-7} {\rm ~kpc~m^{-20/3}}$) is obtained for a scattering screen at an effective $z_L$ of 0.19. The dashed vertical line represents the assumed redshift of the host galaxy of 0.53 (see Section \ref{subsec:Scattering_IGM})}.
    \label{fig:SM_limit_z}
\end{figure}

If the scattering screen is indeed located in the IGM, Eqn. \ref{eqn:tau-SM} allows us to put an upper limit on the strength of the turbulence in the ionised plasma present in the IGM. 
Using the scattering timescale measured for FRB20191107B, and for a screen located midway between the source and the observer, we derive an upper limit on the SM of the scattering-screen in the IGM as ${\rm SM_{IGM} < 8.4 \times 10^{-7}~ kpc~m^{-20/3}}$. 
Since the dependence of scattering on geometry strongly favours plasma located roughly mid-way along the propagation path to the FRB, this limit is a strong function of the location of the scattering-screen and increases sharply as the screen gets closer to the source or the observer. This dependence of the derived upper limit on the SM is shown in Fig \ref{fig:SM_limit_z}.
We have assumed a host redshift of 0.53 in this calculation, and, lowering the host redshift causes the upper limit to relax, albeit gradually. The SM upper-limits for assumed host redshifts of 0.5, 0.4 and 0.3 are $8.5~\times10^{-7}, 9.1~\times~10^{-7}$, and $1.0~\times~10^{-6}~{\rm kpc~m^{-20/3}}$ respectively.

The strongest existing constraints on the strength of turbulence in the IGM come from measurements of the angular broadening of compact extragalactic radio sources like GRB afterglows and AGNs. \cite{Koay2012_AGN_scattering_limit} used multi-frequency observations of sample of 128 compact radio sources and found no evidence for detectible scattering in the IGM for sources in the redshift range $0 < z < 4$.
Towards the most compact $\sim 10~\mu $as sources in their sample, they report an upper limit on angular broadening of $\lesssim 8~\mu$as, constraining the turbulence in the IGM to ${\rm SM \lesssim 3.3 \times 10^{-5} ~kpc~m^{-20/3}}$.
The limit we obtain from scattering timescale measurements of FRB20191107B improves on their limit by more than an order of magnitude, providing the strongest constraints on the strength of turbulence in the IGM so far.

\subsection{Dispersion-scattering relation}
\label{subsec:DM-tau relation}

\begin{figure}
    \centering
    \includegraphics[width=0.45\textwidth]{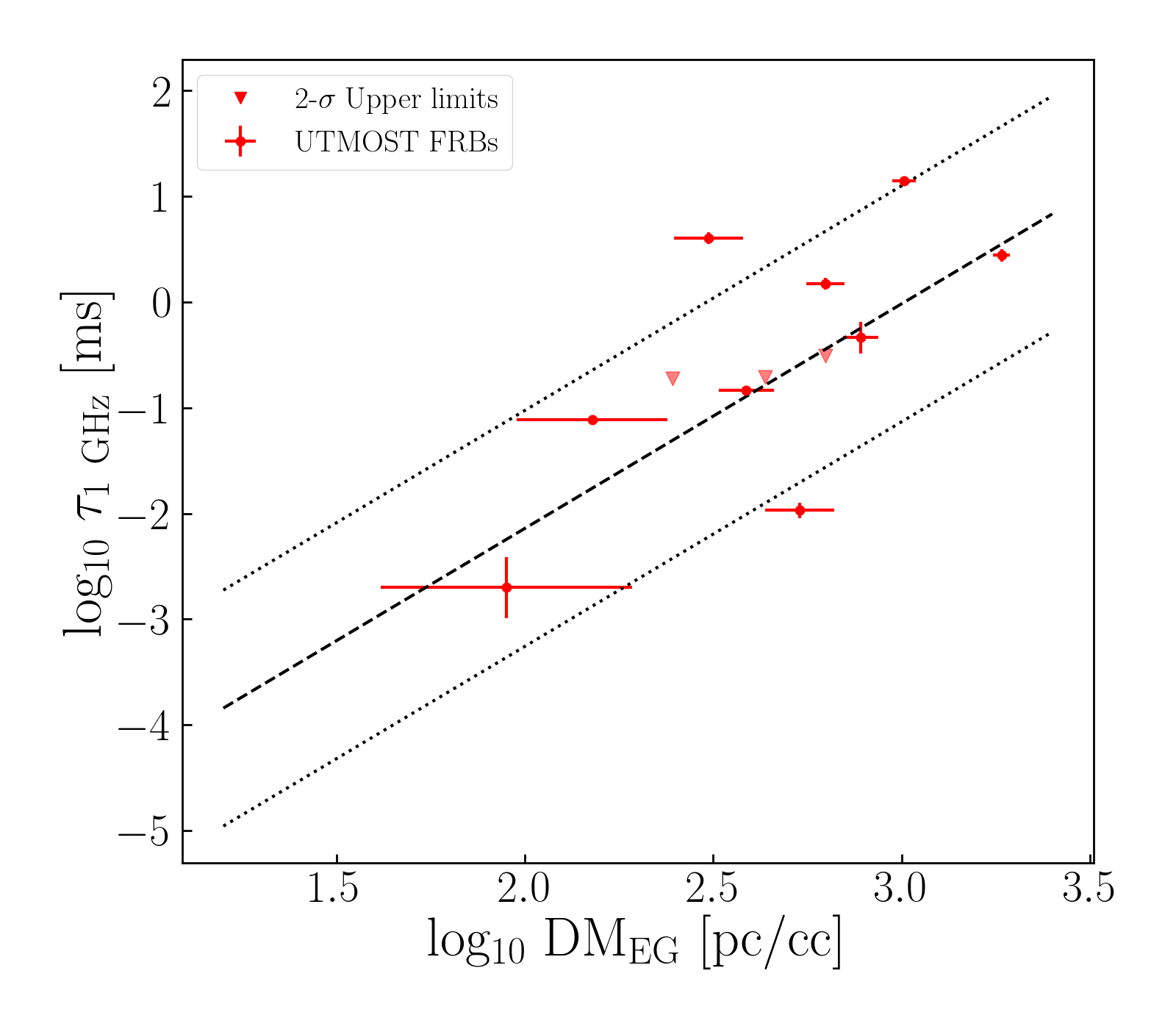}
    \caption{Measurements of the scattering timescales scaled to 1 GHz ($\tau_{\rm 1 GHz}$) plotted against the extragalactic component of the DM (DM$_{\rm EG}$) for the FRBs detected with UTMOST. Downward facing arrows indicate the measurements of 2-$\sigma$ upper limits. The black dashed line shows the best-fit power-law model and the dotted lines indicate the region of $\pm$ 1-$\sigma$ scatter in the fit. For each data point the error in the measurement of scattering timescale is estimated from 1-$\sigma$ scatter in the posterior distribution of the fit model, while the error in the DM$_{\rm EG}$ is taken to be equal to half of the Milky Way contribution (DM$_{MW}$) to FRB's DM budget.}
    \label{fig:DM-tau_UTMOST}
\end{figure}

\begin{figure}
    \centering
    \includegraphics[width=0.45\textwidth]{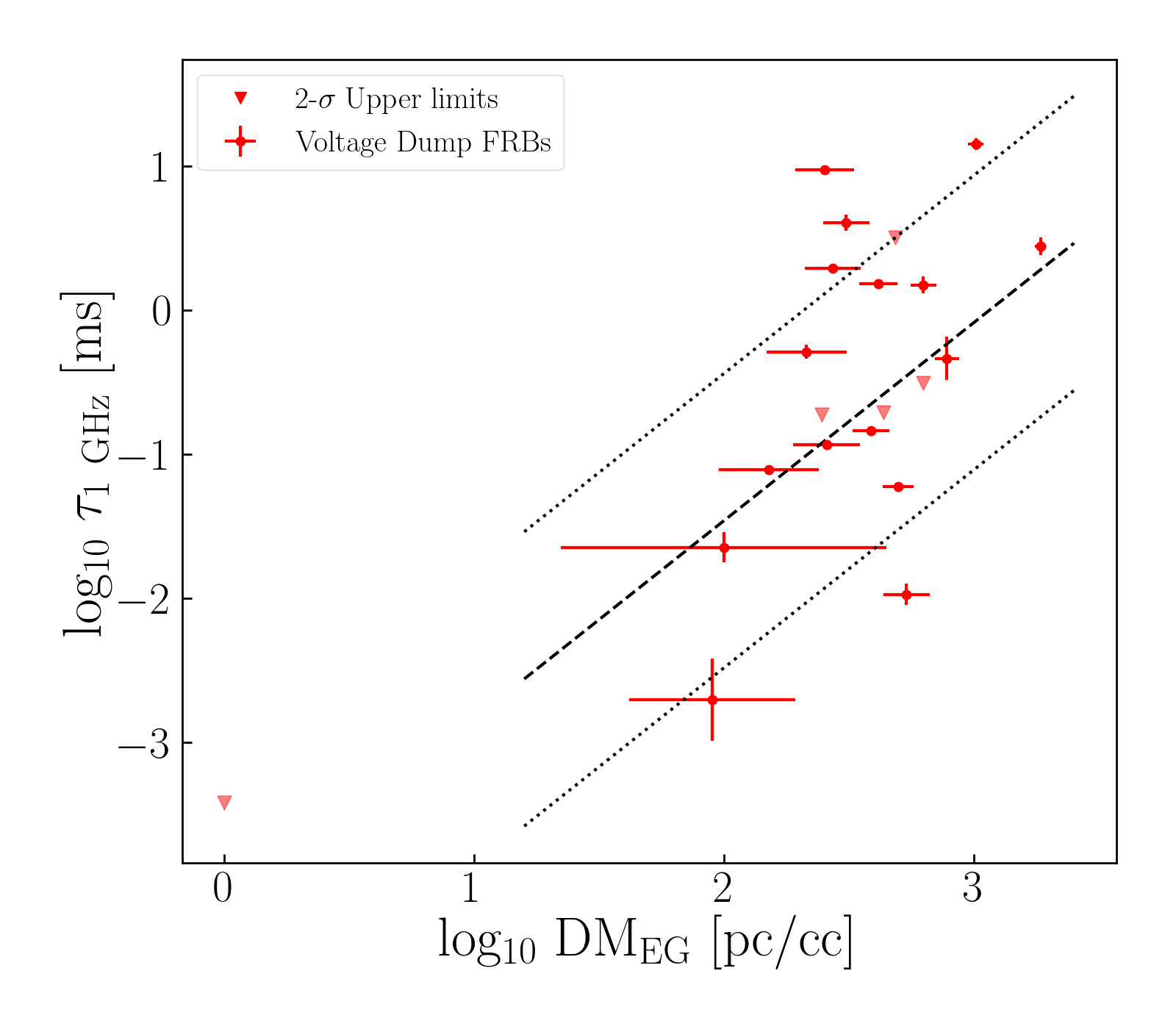}
    \caption{Measurements of the scattering timescales scaled to 1 GHz ($\tau_{\rm 1 GHz}$) plotted against the extragalactic component of the DM (DM$_{\rm EG}$) for the sample of FRBs for which scattering timescales have been measured after coherent dedispersion using the voltage data. Downward facing arrows indicate the measurements of 2-$\sigma$ upper limits. The black dashed line shows the best-fit power-law model and the dotted lines indicate the region of $\pm$ 1-$\sigma$ scatter in the fit. For each data point the error in the measurement of scattering timescale is estimated from 1-$\sigma$ scatter in the posterior distribution of the fit model, while the error in the DM$_{\rm EG}$ is taken to be equal to half of the Milky Way contribution (DM$_{MW}$) to FRB's DM budget.}
    
    \label{fig:DM-tau_voltage}
\end{figure}

\begin{table*}

\centering
\begin{tabular}[c]{|cccccccc|}
\hline
\hline
FRB name & $Gl$($^{\circ}$) & $Gb$($^{\circ}$) & $\tau$ (ms) & DM (\dm) & DM$_{\rm MW}$ (\dm) & $\tau_{\rm MW}$ (\us) & Ref \\
\hline
FRB20170827A & 303.29 & $-51.58$ & 0.00199 & 176.4 & 37.0 & 0.13 & \cite{Farah2018}\\
FRB20170922A & 45.07 & $-38.70$ & 14.14617 & 1111.00 & 45.00 & 0.22 & \cite{Farah2019}\\
FRB20180528A & 258.87 & $-22.35$ & 0.46182 & 899.30 & 70.00 & 0.52 & \cite{Farah2019}\\
FRB20181016A & 345.51 & 22.66 & 2.77090 & 1982.80 & 89.00 & 1.09 & \cite{Farah2019}\\
FRB20181017C & 50.06 & $-46.88$ & 0.07778 & 240.00 & 39.00 & 0.15 & \cite{Farah2019}\\
FRB20181228D & 253.35 & $-26.15$ & <0.18959 & 354.20 & 58.00 & 0.35 & \cite{Farah2019}\\
FRB20190322D & 278.17 & $-36.92$ & <0.31306 & 724.20 & 47.10 & 0.22 & Gupta et al. in prep.\\
FRB20190806B & 89.92 & $-67.25$ & 4.01829 & 388.50 & 30.80 & 0.09 & Gupta et al. in prep.\\
FRB20191107B & 233.40 & 8.83 & 0.01069 & 714.30 & 127.20 & 1.07 & This work\\
FRB20191223B & 278.17 & $-36.92$ & 1.49726 & 724.20 & 47.10 & 0.22 & Gupta et al. in prep.\\
FRB20200508A & 282.02 & $-12.56$ & <0.19445 & 629.00 & 144.90 & 1.32 & Gupta et al. in prep.\\
FRB20200607A & 325.36 & 55.54 & 0.14584 & 466.90 & 30.00 & 0.11 & Gupta et al. in prep.\\
FRB20180924B & 0.74 & $-49.41$ & 1.94215 & 362.40 & 40.50 & 0.16 & \cite{Hao_qiu_2020_ASKAP_FRB_scattering_props}\\
FRB20181112A & 342.60 & $-47.70$ & 0.05998 & 588.80 & 41.70 & 0.17 & \cite{Hao_qiu_2020_ASKAP_FRB_scattering_props}\\
FRB20190102C & 312.65 & $-33.49$ & 0.11710 & 364.40 & 57.40 & 0.33 & \cite{Hao_qiu_2020_ASKAP_FRB_scattering_props}\\
FRB20190608B & 53.21 & $-48.53$ & 9.42513 & 339.50 & 37.30 & 0.14 & \cite{Hao_qiu_2020_ASKAP_FRB_scattering_props}\\
FRB20190611B & 312.94 & $-33.28$ & 0.51410 & 321.40 & 57.80 & 0.34 & \cite{Hao_qiu_2020_ASKAP_FRB_scattering_props}\\
FRB20190711A & 310.91 & $-33.90$ & <3.19883 & 590.50 & 56.50 & 0.32 & \cite{Hao_qiu_2020_ASKAP_FRB_scattering_props}\\
FRB20191001A & 341.23 & $-44.90$ & 1.52133 & 507.90 & 44.00 & 0.19 & \cite{Bhandari2020_frb191001_frb_afterglow}\\
FRB20180916B & 129.71 & $3.73$ & 0.02255 & 348.70 & 199.00 & 4.85 & \cite{Marcote_2020_180916_localisation_Nature}\\
FRB20200120E & 142.20 & $41.22$ & <0.00038 & 87.75 & 41.62 & 0.17 & \cite{Nimmo2021_nanosecond_FRB200120E}\\
\hline
\end{tabular}

\caption{
List of FRB properties ($Gl$, $Gb$, scattering timescales and DM) for the sample of FRBs which have their scattering properties measured after coherent dedispersion.
For comparison, we also list the predicted DM and scattering contribution from the Milky Way ISM along their lines of sight using the NE2001 model. 
This sample has been used in Section \ref{subsec:DM-tau relation} to test the existence of a DM-$\tau$ relation. }
\label{table:voltage_frbs}
\end{table*}

If the IGM is the dominant source of the observed scattering in FRBs, then because SM is the integral of the amplitude of turbulence along the propagation path of an FRB, we expect the effective scattering measure along the line of sight to be correlated with the distance to the source. The plasma in the IGM is already known to be the dominant contributor to the DM budget of the FRBs observed from cosmological distances \citep{Macquart2020_IGM_Baryons_DM_z_relation}.
We therefore expect the existence of DM$-$SM or a DM$-\tau$ relation once a sufficiently large enough sample of FRBs are available.

UTMOST has detected 17 new FRBs so far, of which 12 have voltage data at the native resolution of the telescope and with full phase information available -- the largest sample detected at any telescope. This enables the use of coherent dedispersion to remove instrumental smearing effects and allows careful modelling and analysis of the scattering timescales exhibited by FRBs. 

Using the modelling procedure outlined in Section \ref{subsec:modelling_191107}, we fit scattered Gaussian template profiles to all the FRBs for which we have voltage data available (\citealt{Farah2019}, Gupta et al. in prep).
The modelled values of the scattering-timescale are plotted against the extra-galactic component of the DM of all the voltage-capture FRBs and are shown in Fig. \ref{fig:DM-tau_UTMOST}. For those FRBs which do not show evidence of an exponential scattering tail we plot the 2$-\sigma$ upper limits on the derived values of $\tau$. The values are scaled to 1 GHz assuming a spectral power-law function with an index $-4$: $\tau(\nu) \propto \nu^{-4}$.

We fit the data with a simple power-law model of the form: 
\begin{equation}
    \label{eqn:slopy_line_model}
    \tau_{\rm 1\,GHz} = 10^b \times {\rm DM}_{\rm EG}^m
\end{equation}
where $m$ is the power-law index, $b$ is the scaling parameter. 
For parameter estimation, we use a joint-likelihood function $\mathcal{L}(\theta ~|~ \tau, \tau^{UL})  = \mathcal{L}_1(\theta ~|~ \tau) \times \mathcal{L}_2(\theta ~|~ \tau^{UL})$, where $\mathcal{L}_1$ is the likelihood of the model $M$ with parameters $\theta$ in the presence of scattering timescale measurements $\tau$, and $\mathcal{L}_2$ is the likelihood of the model in the presence of scattering upper-limits $\tau^{UL}$.

We define $\mathcal{L}_1$ as a Gaussian likelihood function of the form (see \citealt{Shannon_and_Cordes_2010_spin_noise_modelling}):
\begin{equation}
    \mathcal{L}_1 (\theta ~|~ \tau) = \prod_{i}^{N} \frac{1}{\sqrt{2\pi\varepsilon_1^2}}{\rm exp} \Big[ - \frac{(\tau_i - M_i(\theta))^2}{2\varepsilon_1^2} \Big],
\end{equation}

\noindent and $\mathcal{L}_2$ as an upper-limit likelihood function of the form:
\begin{equation}
    \mathcal{L}_2(\theta ~|~ \tau^{UL}) = \prod_i^N 1 - \frac{1}{2}\textrm{erfc} \Big[\frac{\tau_i^{UL} - M_i(\theta)}{\sqrt{2}\varepsilon_2}\Big] 
\end{equation}

\noindent $\varepsilon_1$ and $\varepsilon_2$ quantify the uncertainty in the residual, and can be calculated as the quadrature sum of individual uncertainties using the following relations:
\begin{equation}
    \varepsilon_1 = \sqrt{m^2 (\Delta\log {\rm DM}_{EG})^2 + (\Delta\log\tau)^2 + s^2}
\end{equation}
\begin{equation}
    \varepsilon_2 = \sqrt{m^2 (\Delta\log {\rm DM}_{EG})^2 + s^2}
\end{equation}

\noindent where $\Delta$DM$_{\rm EG}$ is the error in the estimated value of DM$_{EG}$, $\Delta \tau$ is the error in the measured value of $\tau$, and $s$ is an additional parameter introduced to quantify the scatter in the best fit model. 
 We use the DYNESTY nested sampler \citep{dynesty_sampler2020} to obtain the Bayesian posteriors and evidences of our models.
 $\Delta \tau$ are computed using a 1-$\sigma$ confidence interval in the posterior distributions of the best fit templates to the FRB profiles, whereas we take half of the predicted DM contribution of the Milky Way's ISM (NE2001 model; \citealt{NE2001}) as the error in the estimated value of DM$_{\rm EG}$.

We find that the data are well fit by the model with parameters $m=2^{+1}_{-1}, b=-6^{+3}_{-3}$ and a scatter $s=1.1^{+0.4}_{-0.3}$. The best fit model, along with the measured scatter, is plotted in Fig \ref{fig:DM-tau_UTMOST}.

To test the hypothesis that the data support this power-law relation, we compute the Bayes factor ($\mathcal{B}$) using the ratio of the marginal likelihood of our power-law fit to the marginal likelihood of the fit of a model with a power-law index of $m=0$, i.e. a model with no correlation between the observed $\tau$ and the DM$_{EG}$ of the FRBs. We adopt the Jeffreys scale \citep{jeffreys1998theory_bayes_factor_citation, Trotta_2008_Bayes_factor_Jeffreys_scale} for the interpretation of the evaluated Bayes factor, and find that our data provides negligible or only marginal evidence in favour of the power-law model fit ($m \neq 0$), with $\log \mathcal{B} < 1$. 

We expand our sample of bursts by adding all those FRBs reported in the literature whose scattering timescales have been measured accurately using coherently dedispersed data. The names and scattering timescales of all FRBs used in this sample are listed in Table \ref{table:voltage_frbs}.

Using this expanded sample of FRBs, the values of the parameters of the power-law model that best fits the data are: $m=1.4^{+0.7}_{-0.6}, b=-4.2^{+1.7}_{-1.8}$, and a scatter $s=1.0^{+0.2}_{-0.2}$. Once again, we also fit the data with a power-law index fixed at 0 indicating no correlation between the observed $\tau$ and DM$_{EG}$, and compare the marginal likelihoods of the two fits in order to evaluate the Bayes factor $\mathcal{B}$. We find that our sample shows slightly increased support for the power-law model, however, the evidence is still weak, with $1 < \log \mathcal{B} < 2.5$.

Therefore, we conclude that our data hints at a potential DM-$\tau$ relation in FRBs, however, we do not find strong evidence to establish the existence of such a relation. 
This result is consistent with the findings of \cite{Ravi2019_observed_prop}, \cite{Hao_qiu_2020_ASKAP_FRB_scattering_props} and \cite{First_CHIME_catalog_2021} who investigated the scattering properties of the sample of FRBs detected at Parkes, ASKAP, and CHIME respectively, and found no evidence for a potential DM-$\tau$ relation in their samples of FRBs. 

However, their sample of FRBs were not coherently dedispersed and had relatively coarser time resolution data available than the sample analysed in this work. Therefore, it is possible that true scattering timescales of FRBs in their samples were biased towards larger values due to poorer frequency and time resolution of their data, resulting in diluting of the evidence for the existence of the DM-$\tau$ relation.

Future surveys hold the promise of detecting large numbers of FRBs with coherently dedispersed data which can provide a large sample of accurately measured scattering timescales and DM, and will be instrumental in establishing or excluding the existence of a DM-$\tau$ relation in FRBs. 

In addition, it is important to note that while past surveys like those at CHIME, Parkes and the ASKAP telescopes have measurement biases against FRBs with widths less than their sampling time due to the lack of access to raw voltage data, surveys at telescopes like UTMOST which have access to coherently dedispersed data of majority of their FRBs still suffer from detection bias against FRBs with large total widths \citep{Gupta_et_al_2021_mock_frb_injection_utmost, Connor2019_interpreting_the_distribution_of_frb_observables}.
Future modelling of the DM-$\tau$ relation will require careful investigation and correction for these detection and measurement biases in the detected sample of FRBs by a given telescope.

\section{On the origin of scattering in FRBs}
\label{sec: Origin of scattering}

Similarly to pulsars, FRBs as a population exhibit a wide range of scattering times, spanning several orders of magnitude, and as with pulsars, scattering is likely to arise in turbulent ionised media located along the line-of-sight. In this section we examine what can be inferred about the location of the scattering medium from the properties of our UTMOST FRBs. We consider all likely causes of scattering along the line-of-sight, from the ISM and halo of the Milky Way, the IGM, the FRB's host galaxy, and the ``circumburst'' environment in the immediate vicinity of the FRB.  

\subsection{ISM of the Milky Way}
\label{subsec: origin_of_scattering:ISM of Milky Way}

Estimates of scattering timescale along a given line of sight due to the ISM of the Milky Way are available from the NE2001 and/or YMW16 models \citep{NE2001, YMW16}, which can be compared against the measured scattering timescale for FRBs. We list these predicted scattering timescales from the NE2001 model, along with the measured scattering timescales for FRBs in our sample in Table \ref{table:voltage_frbs} for side-by-side comparison.
Most FRBs in our sample have been detected at high Galactic latitudes, where the estimates of scattering from the NE2001 and YMW16 models, due to the ISM, are several orders of magnitude smaller than the observed FRB scattering timescales. \cite{Ocker_and_Cordes_2021_Halo_scattering_limits} have also modelled the contribution of the thick disc of the Milky Way ISM, and they predict scattering timescales in the range 29 ns to 0.25\us\ for lines of sight sampling the thick disc above galactic latitudes $>20^{\rm o}$, also making a negligible contribution to the observed scattering timescales in FRBs.




\subsection{The Milky Way halo}
\label{subsec: origin_of_scattering:halo of Milky Way}

The density profile of the tenuous ionised material in the Milky Way halo remains poorly constrained by observations, due to small numbers of suitable tracers and its very low emissivity \citep{Gupta_et_al_2012_MW_Halo_EM}. FRBs have opened up an entirely new means of probing the density and turbulence of this material \citep{Platts2020_MW_halo}. \cite{Ocker_and_Cordes_2021_Halo_scattering_limits} have used two nearby repeating FRBs, namely, FRB20121102A and FRB20180916B, to constrain the amplitude of scattering due to the Milky Way halo. They set an upper limit on scattering timescale at 1 GHz of $\lesssim$ 12 $\mu$s, which, as they point out, is comparable to scattering effects of the Galactic disc ISM at high Galactic latitudes. This rules out the Milky Way halo as the origin of the scattering timescales observed in our FRBs.

\subsection{The Intergalactic Medium}
\label{subsec: origin_of_scattering:IGM}

In the previous section (Section \ref{subsec:DM-tau relation}) we have shown that there is little to no evidence for the existence of a strong DM-$\tau$ relation amongst FRBs (Section \ref{subsec:DM-tau relation}). Additionally, the stringent upper limit we set on the strength (SM $< 8.4 \times 10^{-7} {\rm ~kpc~m^{-20/3}}$) (see Fig \ref{fig:SM_limit_z}) of turbulence in the IGM using FRB20191107B suggests that the diffuse ionised plasma in the IGM is unlikely to be the dominant source of scattering observed in FRBs. The rest of our FRBs have scattering timescales much greater than this turbulence can provide. 

Gaseous disks and the circumgalactic medium (CGM) of intervening galaxies can intersect the lines of sight of FRBs as the radiation traverses the IGM, as has been seen in FRBs reported by \citep{Ravi2019_localisation, Prochaska2019_FRB181112, Simha2020_190608_IGM_cosmic_web, Connor2020_FBR191108}. 

The probability of intersecting a galaxy disc is expected to be low and is computed to be only approximately 5\% \citep{Macquart&Koay2013}) for a sources up to $z < 1.5$. While true that, due to the geometric lever arm effect, the ISM of foreground galaxies would dominate the observed scattering along that line of sight, the low probability of intersecting such foreground systems means that intervening galactic discs can only be used to account for the observed scattering in at most one or two FRBs in our sample. 

\cite{Vedantham2019_scattering_CGM} have argued that the CGM of intervening galaxies can contribute between 0.1-10 ms of temporal broadening (at 30 cm wavelength) due to scattering. 
However, \cite{Cho2020_pfbinverted_181112} have set an upper limit on the scattering timescale of 20 $\mu$s for FRB20181112A, despite its line of sight passing through the halo of a foreground galaxy, showing that even smaller scattering times than those estimated by \cite{Vedantham2019_scattering_CGM} are possible. 
Extending the analysis of \cite{Cho2020_pfbinverted_181112}, \cite{Simha2020_190608_IGM_cosmic_web} studied the  of the foreground halo present along the line of sight to FRB20190608B and find that the ionised plasma in the halo of the intervening galaxy cannot produce sufficient scattering to account for the observed scattering timescale in the FRB. 
\cite{Ocker_and_Cordes_2021_Halo_scattering_limits} also use the scattering timescales observed in FRB20191108A and repeat bursts from FRB20190816B to constrain the density of ionised plasma in the galaxy halos, and conclude that the halos may not be sufficiently turbulent and very little scattering occurs in the intervening galaxy halos along these FRB lines of sight.

Therefore, for most of the FRBs in our sample, we are able to rule out all sources of scattering along the line of sight, up to the host galaxy itself, where the ISM and/or the circumburst medium could explain the observed scattering properties in our FRBs.

\subsection{Scattering in the FRB host galaxy}

\label{subsec: origin_of_scattering:host galaxy}


Circumstantial evidence for scattering in the host galaxies of FRBs has been reported in the literature. \cite{Farah2018} and \cite{Masui2015} found evidence for the presence of two scattering screens along the lines of sight to FRB20170827A and FRB20110523A respectively, and suggest that the spectral modulations associated with the temporal broadening of the bursts can be attributed to a scattering screen located within the host galaxy of the burst, while the broader scale spectral modulations are consistent with originating from within the Milky Way.

We next investigate properties of the host galaxy ISM, under the assumption that it is the dominant source of scattering observed in FRBs. To enable a direct comparison with the Milky Way's ISM, we scale the observed scattering timescales of FRBs to take out the effect of the expansion of the Universe with redshift.

The exponential decay timescale of a signal due to scattering from a turbulent medium with power-law variation in densities (approximated as a single screen) scale with frequency ($\nu$) as a power-law: $\tau(\nu) \propto \nu^{-\alpha}$, where $\alpha$ is found to have the value between 4.0 (for square power-law variation distribution) and 4.4 (for Kolmogorov distribution) \citep{Lee_and_Jokiipi1976, Oswald21_Meerkat_pulsar_scattering_indices}. 

At Earth, the frequency of emission has been redshifted and the scattering time dilated by a factor $1+z$. Consequently, the scattering produced in the host galaxy ($\tau_{\rm host}$) is modified by a combination of two effects:
\begin{equation}
    \tau_{\rm obs} = \tau_{\rm host} \frac{(1+z)}{~~(1+z)^{\alpha}}.
\end{equation}
For $\alpha = 4$, the scattering thus scales as $(1+z)^{-3}$.

In addition to the cosmological effects of frequency redshift and time dilation, the change in the relative distance to the screen with respect to the observer and the source needs to be considered when estimating $\tau_{\rm host}$ from $\tau_{\rm obs}$.
\cite{Cordes2016_FRB_scattering} have shown that the observed scattering within the host galaxy would be a factor of $\sim$3 lower than that observed from the Earth, as the waves would be planar when they arrive at the screen from a distant galaxy (after invoking reciprocity) as opposed to spherical when observed from within the host galaxy.

\begin{figure}
    \centering
    \includegraphics[width=0.45\textwidth]{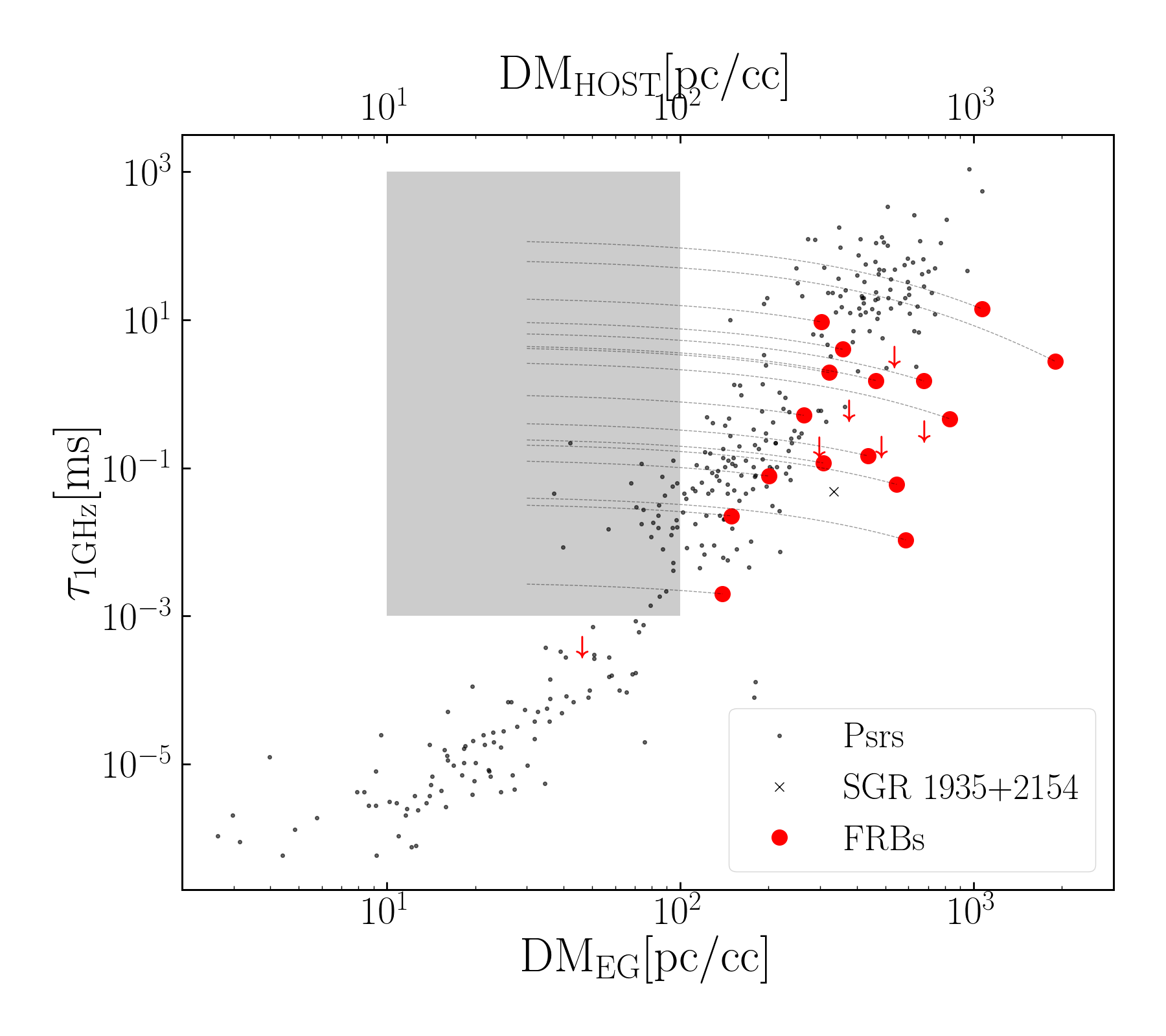}
    \caption{Expected scattering timescales of FRBs after scaling to the frequency of 1 GHz in the rest frame of the host galaxy under the assumption that the host galaxy is the dominant source of scattering. The red circles show the measured values of scattering timescale at Earth. Taking out the effect of redshifting of frequency and dilation of time, the scattering timescales scale along the dashed grey lines (following Eqn. \ref{eqn: tau_host_from_tau_obs}) for each FRB and would lie in the region highlighted in the grey box depending upon the DM contributed by the host galaxy.}
    \label{fig:DM_tau_burst}
\end{figure}

Therefore, under the assumption that the host's ISM is the dominant source of scattering, $\tau_{\rm host}$ can be related to $\tau_{\rm obs}$ as:
\begin{equation}
\tau_\mathrm{host} = 1/3\times \tau_\mathrm{obs}~(1+z)^{3}.
\label{eqn: tau_host_from_tau_obs}
\end{equation}
We scale the observed scattering timescale of the FRBs in our sample according to the above relation, and plot them over the observed scattering timescales of Milky Way pulsars in Fig \ref{fig:DM_tau_burst}. 

The host galaxy contribution to the total DM is still not well constrained and a value of $\lesssim 100$ \dm\ has been commonly used in the literature. Assuming that DM$_{\rm host}$ typically lies in the range 10 to 100 \dm, we identify a region in the $\tau$-DM$_{\rm host}$ space where our scaled scattering timescales of FRBs would lie. This region is shown as the grey box in Fig \ref{fig:DM_tau_burst}.

It is evident that the values of $\tau_{\rm host}$ are orders of magnitude larger than the scattering timescales observed in the Milky Way pulsars in the same DM range as adopted for the the hosts (DM$_{\rm host}$). Similar levels of scattering are observed from those Milky Way pulsars which show much larger values of DM ($>100$ \dm), and typically lie in the dense galactic disc of our galaxy.
Since scattering in pulsars is dominated by the ISM of the Milky Way, associating the scattering in FRBs to the ISM of their host galaxies requires that the ISM of the host galaxies is many orders of magnitude more turbulent than the ISM of the Milky Way (see \citealt{Xu_and_Zhang_2016_origin_of_scattering}). FRBs would have to occur predominantly in galaxies with highly turbulent conditions in the ISM.

Alternatively, the FRB lines of sight may be traversing larger distances through the host galaxy's ISM, increasing the likelihood of encountering multiple turbulent clumps along their path and resulting in large values of scattering similar to those seen in high-DM pulsars in the Milky Way. However, this would require that the host galaxies also produce large contributions ($>100$ \dm) to the observed DM of the FRB, and in case of spiral galaxies, that they appear edge-on at high inclinations when observed from the Earth.

In contrast, the small sample of FRB host galaxies that have been identified so far have shown a wide variety of properties, such as a spread of $>2$ orders of magnitude in the star formation rate and total stellar mass \citep{Heintz_2020_host_galaxies_ASKAP, Mannings_2020_host_galaxies_env}. These values do not indicate that FRB progenitors reside exclusively in galaxies which are likely to have highly turbulent ISM (expected in galaxies with increased energy feedback from star formation \cite{Xu_and_Zhang_2016_origin_of_scattering}). Furthermore, the majority of the localised FRBs have been found to originate in the outskirts of their host galaxies (the FRBs are offset physically in the range $0.4-5.3 R_\mathrm{eff}$, where $R_\mathrm{eff}$ is the host galaxy's effective radius: \cite{Heintz_2020_host_galaxies_ASKAP, Mannings_2020_host_galaxies_env}) and the inclinations have been found to be generally low. 
\cite{Niino_2020_host_galaxy_DM} analyzed distributions of DM of the FRB samples observed by ASKAP and the Parkes radio telescope and estimated the total DM contribution of the host galaxies (including the host's halo) to be $\sim120$ \dm. 
More recently, \cite{Chittidi2020_190608_host_galaxy_disection} estimated the DM contribution from the host galaxy of FRB20190608B to be $\approx 85$ \dm, and the scattering contribution to be only $3\mu$s out of the 3.3 ms scattering timescale measured at 1.3 GHz by \cite{Day_et_al_MNRAS}. 

We conclude that the properties of the localised sample of host galaxies do not support the scenario that the ISM of the host galaxy is the dominant source of scattering in most FRBs.

This leaves the turbulence in the circumburst environment of the source as most likely to produce the observed scattering in FRBs, as has been suggested previously by \cite{Masui2015, Spitler2016}.
The large scatter in the scattering timescales of FRBs as a group would appear to be better explained by invoking a diversity in the circumburst medium of different sources, as opposed to invoking a wide range of turbulence in the host ISM or in the IGM.

We make an order-of-magnitude estimate for scattering due to the circumburst medium as follows: assuming a fiducial distance of 1 pc to the scattering screen located in the circumburst medium, then using Eqn \ref{eqn:tau-SM} we derive SM $= 7\times10^{20} {\rm m^{-17/3}}$ (or $\sim 24~ {\rm kpc~m^{-20/3}}$) for the environment of FRB20191107B, but would be up to 4 orders of magnitude larger for some of the most scattered FRBs in our sample. This is much larger than the typical values of SM observed for Milky Way pulsars, but is comparable to the SM values of some of the overdense regions identified by \cite{NE2001_II_2003} along the lines of sight to some pulsars in the Milky Way. For example, the Vela supernova remnant has an estimated SM of $\sim34$  kpc m$^{-20/3}$, while some lines of sight to a few extragalactic sources (e.g. NGC6334B) have been estimated to have even higher values of SM \citep{NE2001_II_2003}.
Consequently, the high values of SM required for the turbulent clumps of ionised plasma present along the line of sight to FRBs in the host galaxies do have some analogue in the Milky Way. 
The presence of such dense and turbulent media in the vicinity of the source can also produce a notable contribution to the Rotation Measure (RM) of an FRB, and can lead to a RM-$\tau$ relation in FRBs. Since UTMOST uses only a single circular polarisation to observe and detect FRBs \citep{utmost}, we do not have the RM values of the majority of FRBs in our sample. 
As more RM and $\tau$ values for FRBs are reported from large ongoing surveys, searching for a potential RM-$\tau$ relation in FRBs can shed more light on the origin of scattering in the vicinity of the burst progenitor.

\section{Conclusions}
\label{sec: Conclusions}

We report the detection of FRB20191107B with UTMOST. Using the raw voltage data captured for the FRB we have analysed the temporal properties of FRB20191107B - the narrowest FRB so far detected with UTMOST, at the native instrument time resolution (10.24 $\mu$s). We model three components in the burst profile and measure a DM of 715.7 \dm, scattering time $\tau = 21.4 \mu$s, and an intrinsic width of only 11.3 $\mu$s. 

We model the limitation in the sensitivity for narrow width FRBs due to the limited time and frequency resolution of a survey. Assuming a power-law distribution of burst energies with a power-law index of $-1.8$, we find that UTMOST's FRB survey would only detect $\sim5$\% of FRBs at the measured width of FRB20191107B.
Using the reported sensitivities of other prominent radio telescopes surveys like CHIME, Murriyang (also known as the Parkes radio telescope) and ASKAP, we find that most current FRB surveys are also similarly insensitive to FRBs with intrinsically narrow widths.

The detection of a single event --- FRB20191107B suggests that a significant population of such FRBs may exist which evades the searched parameter space of most active FRB surveys. 
For FRBs originating at a given redshift, FRBs with narrower intrinsic widths will have a higher flux, and are more likely to be detected than the FRBs with wide widths. Therefore, improving the sensitivity of a survey to narrow width events will also make the survey more sensitive to high-redshift FRBs.

FRBs carry with them the signature of the properties of the ionised plasma that lies along their propagation path. FRBs coming from large Gpc distances offer unique probes into the properties of their host galaxies, any intervening galaxies and the IGM along their lines of sight \citep{Ravi_et_al2019_tomography_of_universe}. Detecting a large sample ($10^3$-$10^5$) of FRBs from high redshifts ($z > 3$) has been deemed necessary to enable their use as cosmological probes into the history of the evolution of Universe \citep{Caleb2019_EoR, Fialkov_and_Loeb2016_EoR, Pagano_et_al2021_EoR, Hashimoto_et_al2021EoR}. 
Our findings suggest that increasing the sensitivity for narrow width FRBs would be useful in probing the population of high redshift FRBs and future surveys should be designed while considering the benefits and costs of searching for FRBs at a higher time and frequency resolution.

We use the observed scattering timescale of FRB20191107B to place a stringent upper limit on strength of turbulence in the IGM (SM $< 8.4 \times 10^{-7} {\rm ~kpc~m^{-20/3}}$ (see Section \ref{subsec:DM-tau relation}, Fig \ref{fig:SM_limit_z}).
We build a sample of 21 FRBs for which scattering timescale measurements have been reported using analysis of the high-time resolution data with full phase information retained via voltage capture and search for a DM$_{EG}-\tau$ relation in FRBs but find only marginal evidence in favour of its existence.
The lack of evidence in favour of a DM$_{EG}-\tau$ relation and the strong upper limit on the strength of turbulence in the IGM along the line of sight to FRB20191107B argue against IGM being the dominant source of scattering in FRBs.

Recent detections of microsecond scale scattering in the local population of repeating FRBs $-$ FRB20180916B and FRB20200120E \citep{Marcote_2020_180916_localisation_Nature, Nimmo2021_nanosecond_FRB200120E} have been used by \cite{Ocker_and_Cordes_2021_Halo_scattering_limits} to strongly limit the amount of scattering produced by the Milky Way halo ($<12 \mu$s). 
We identify the circumburst medium of the FRB progenitor as the most likely origin of scattering in most FRBs, and compare the required levels of turbulence in the circumburst medium with some of the dense ionised regions found in Milky Way. 
We find that the circumburst environment of FRBs must be much more turbulent than the environment of an average Milky Way pulsar, which is consistent with the recent findings of \cite{Chawla2021_scattering_CHIME_FRBs}.

\section*{Acknowledgements}
The authors would like to thank Ryan M.~Shannon and Stefan Os{\l}owski for important discussions during the preparation of this paper. The Molonglo Observatory is owned and operated by the University of Sydney, with support from the School of Physics and the University. The UTMOST project is also supported by the Swinburne University of Technology. We acknowledge the Australian Research Council grants CE110001020 (CAASTRO) and the Laureate Fellowship FL150100148. ATD is supported by an ARC Future Fellowship grant FT150100415. This research made use of numpy \citep{numpy}, pandas \citep{pandas-official}, matplotlib \citep{matplotlib}, astropy \cite{astropy:2013, astropy:2018}, BILBY \citep{Bilby} and PSRCHIVE \citep{PSRCHIVE2004} packages.

\section*{Data Availability}

The coherently dedispersed dynamic spectra, the frequency averaged time series and the codes used in modelling of the burst profile are available at the github repository: \url{https://github.com/vivgastro/FRB191107_paper_data_release/}. The codes used to estimate the detection fraction of narrow FRBs in Section \ref{sec:rates of narrow FRBs} are also available in the same repository. Further raw data can be made available upon reasonable request to the authors.



\bibliographystyle{mnras}
\bibliography{bibliography_big, python_packages} 

\begin{thebibliography}{}
\makeatletter
\relax
\def\mn@urlcharsother{\let\do\@makeother \do\$\do\&\do\#\do\^\do\_\do\%\do\~}
\def\mn@doi{\begingroup\mn@urlcharsother \@ifnextchar [ {\mn@doi@}
  {\mn@doi@[]}}
\def\mn@doi@[#1]#2{\def\@tempa{#1}\ifx\@tempa\@empty \href
  {http://dx.doi.org/#2} {doi:#2}\else \href {http://dx.doi.org/#2} {#1}\fi
  \endgroup}
\def\mn@eprint#1#2{\mn@eprint@#1:#2::\@nil}
\def\mn@eprint@arXiv#1{\href {http://arxiv.org/abs/#1} {{\tt arXiv:#1}}}
\def\mn@eprint@dblp#1{\href {http://dblp.uni-trier.de/rec/bibtex/#1.xml}
  {dblp:#1}}
\def\mn@eprint@#1:#2:#3:#4\@nil{\def\@tempa {#1}\def\@tempb {#2}\def\@tempc
  {#3}\ifx \@tempc \@empty \let \@tempc \@tempb \let \@tempb \@tempa \fi \ifx
  \@tempb \@empty \def\@tempb {arXiv}\fi \@ifundefined
  {mn@eprint@\@tempb}{\@tempb:\@tempc}{\expandafter \expandafter \csname
  mn@eprint@\@tempb\endcsname \expandafter{\@tempc}}}

\bibitem[\protect\citeauthoryear{{Ashton} et~al.,}{{Ashton}
  et~al.}{2019}]{Bilby}
{Ashton} G.,  et~al., 2019, \mn@doi [\apjs] {10.3847/1538-4365/ab06fc}, \href
  {https://ui.adsabs.harvard.edu/abs/2019ApJS..241...27A} {241, 27}

\bibitem[\protect\citeauthoryear{{Astropy Collaboration} et~al.,}{{Astropy
  Collaboration} et~al.}{2013}]{astropy:2013}
{Astropy Collaboration} et~al., 2013, \mn@doi [\aap]
  {10.1051/0004-6361/201322068}, \href
  {http://adsabs.harvard.edu/abs/2013A%26A...558A..33A} {558, A33}

\bibitem[\protect\citeauthoryear{{Astropy Collaboration} et~al.,}{{Astropy
  Collaboration} et~al.}{2018}]{astropy:2018}
{Astropy Collaboration} et~al., 2018, \mn@doi [\aj] {10.3847/1538-3881/aabc4f},
  \href {https://ui.adsabs.harvard.edu/abs/2018AJ....156..123A} {156, 123}

\bibitem[\protect\citeauthoryear{{Bailes} et~al.,}{{Bailes}
  et~al.}{2017}]{utmost}
{Bailes} M.,  et~al., 2017, \mn@doi [\pasa] {10.1017/pasa.2017.39}, \href
  {https://ui.adsabs.harvard.edu/abs/2017PASA...34...45B} {34, e045}

\bibitem[\protect\citeauthoryear{{Bhandari} et~al.,}{{Bhandari}
  et~al.}{2020}]{Bhandari2020_frb191001_frb_afterglow}
{Bhandari} S.,  et~al., 2020, \mn@doi [\apjl] {10.3847/2041-8213/abb462}, \href
  {https://ui.adsabs.harvard.edu/abs/2020ApJ...901L..20B} {901, L20}

\bibitem[\protect\citeauthoryear{{Bhardwaj} et~al.,}{{Bhardwaj}
  et~al.}{2021}]{Bhardwaj2021_M81_frb_repeater_frb20200120E}
{Bhardwaj} M.,  et~al., 2021, \mn@doi [\apjl] {10.3847/2041-8213/abeaa6}, \href
  {https://ui.adsabs.harvard.edu/abs/2021ApJ...910L..18B} {910, L18}

\bibitem[\protect\citeauthoryear{{Bochenek}, {Ravi}, {Belov}, {Hallinan},
  {Kocz}, {Kulkarni}  \& {McKenna}}{{Bochenek}
  et~al.}{2020}]{Bochenek2020_STARE2_FRB}
{Bochenek} C.~D.,  {Ravi} V.,  {Belov} K.~V.,  {Hallinan} G.,  {Kocz} J.,
  {Kulkarni} S.~R.,   {McKenna} D.~L.,  2020, arXiv e-prints, \href
  {https://ui.adsabs.harvard.edu/abs/2020arXiv200510828B} {p. arXiv:2005.10828}

\bibitem[\protect\citeauthoryear{{Caleb} et~al.,}{{Caleb}
  et~al.}{2017}]{Caleb_3frbs}
{Caleb} M.,  et~al., 2017, \mn@doi [\mnras] {10.1093/mnras/stx638}, \href
  {https://ui.adsabs.harvard.edu/abs/2017MNRAS.468.3746C} {468, 3746}

\bibitem[\protect\citeauthoryear{{Caleb}, {Flynn}  \& {Stappers}}{{Caleb}
  et~al.}{2019}]{Caleb2019_EoR}
{Caleb} M.,  {Flynn} C.,   {Stappers} B.~W.,  2019, \mn@doi [\mnras]
  {10.1093/mnras/stz571}, \href
  {https://ui.adsabs.harvard.edu/abs/2019MNRAS.485.2281C} {485, 2281}

\bibitem[\protect\citeauthoryear{{Chawla} et~al.,}{{Chawla}
  et~al.}{2021}]{Chawla2021_scattering_CHIME_FRBs}
{Chawla} P.,  et~al., 2021, arXiv e-prints, \href
  {https://ui.adsabs.harvard.edu/abs/2021arXiv210710858C} {p. arXiv:2107.10858}

\bibitem[\protect\citeauthoryear{{Chittidi} et~al.,}{{Chittidi}
  et~al.}{2020}]{Chittidi2020_190608_host_galaxy_disection}
{Chittidi} J.~S.,  et~al., 2020, arXiv e-prints, \href
  {https://ui.adsabs.harvard.edu/abs/2020arXiv200513158C} {p. arXiv:2005.13158}

\bibitem[\protect\citeauthoryear{{Cho} et~al.,}{{Cho}
  et~al.}{2020}]{Cho2020_pfbinverted_181112}
{Cho} H.,  et~al., 2020, \mn@doi [\apjl] {10.3847/2041-8213/ab7824}, \href
  {https://ui.adsabs.harvard.edu/abs/2020ApJ...891L..38C} {891, L38}

\bibitem[\protect\citeauthoryear{{Connor}}{{Connor}}{2019}]{Connor2019_interpreting_the_distribution_of_frb_observables}
{Connor} L.,  2019, \mn@doi [\mnras] {10.1093/mnras/stz1666}, \href
  {https://ui.adsabs.harvard.edu/abs/2019MNRAS.487.5753C} {487, 5753}

\bibitem[\protect\citeauthoryear{{Connor} et~al.,}{{Connor}
  et~al.}{2020}]{Connor2020_FBR191108}
{Connor} L.,  et~al., 2020, arXiv e-prints, \href
  {https://ui.adsabs.harvard.edu/abs/2020arXiv200201399C} {p. arXiv:2002.01399}

\bibitem[\protect\citeauthoryear{{Cordes} \& {Chatterjee}}{{Cordes} \&
  {Chatterjee}}{2019}]{Cordes&Chatterjee2019_FRB_review}
{Cordes} J.~M.,  {Chatterjee} S.,  2019, \mn@doi [\araa]
  {10.1146/annurev-astro-091918-104501}, \href
  {https://ui.adsabs.harvard.edu/abs/2019ARA&A..57..417C} {57, 417}

\bibitem[\protect\citeauthoryear{{Cordes} \& {Lazio}}{{Cordes} \&
  {Lazio}}{1991}]{Cordes_and_Lazio_1991_Scintillation_theory}
{Cordes} J.~M.,  {Lazio} T.~J.,  1991, \mn@doi [\apj] {10.1086/170261}, \href
  {https://ui.adsabs.harvard.edu/abs/1991ApJ...376..123C} {376, 123}

\bibitem[\protect\citeauthoryear{{Cordes} \& {Lazio}}{{Cordes} \&
  {Lazio}}{2002}]{NE2001}
{Cordes} J.~M.,  {Lazio} T.~J.~W.,  2002, arXiv e-prints, \href
  {https://ui.adsabs.harvard.edu/abs/2002astro.ph..7156C} {pp
  astro--ph/0207156}

\bibitem[\protect\citeauthoryear{{Cordes} \& {Lazio}}{{Cordes} \&
  {Lazio}}{2003}]{NE2001_II_2003}
{Cordes} J.~M.,  {Lazio} T.~J.~W.,  2003, arXiv e-prints, \href
  {https://ui.adsabs.harvard.edu/abs/2003astro.ph..1598C} {pp
  astro--ph/0301598}

\bibitem[\protect\citeauthoryear{{Cordes}, {Wharton}, {Spitler}, {Chatterjee}
  \& {Wasserman}}{{Cordes} et~al.}{2016}]{Cordes2016_FRB_scattering}
{Cordes} J.~M.,  {Wharton} R.~S.,  {Spitler} L.~G.,  {Chatterjee} S.,
  {Wasserman} I.,  2016, arXiv e-prints, \href
  {https://ui.adsabs.harvard.edu/abs/2016arXiv160505890C} {p. arXiv:1605.05890}

\bibitem[\protect\citeauthoryear{{Day} et~al.,}{{Day}
  et~al.}{2020}]{Day_et_al_MNRAS}
{Day} C.~K.,  et~al., 2020, \mn@doi [\mnras] {10.1093/mnras/staa2138}, \href
  {https://ui.adsabs.harvard.edu/abs/2020MNRAS.497.3335D} {497, 3335}

\bibitem[\protect\citeauthoryear{{Farah} et~al.,}{{Farah}
  et~al.}{2018}]{Farah2018}
{Farah} W.,  et~al., 2018, \mn@doi [\mnras] {10.1093/mnras/sty1122}, \href
  {https://ui.adsabs.harvard.edu/abs/2018MNRAS.478.1209F} {478, 1209}

\bibitem[\protect\citeauthoryear{{Farah} et~al.,}{{Farah}
  et~al.}{2019}]{Farah2019}
{Farah} W.,  et~al., 2019, \mn@doi [\mnras] {10.1093/mnras/stz1748}, \href
  {https://ui.adsabs.harvard.edu/abs/2019MNRAS.488.2989F} {488, 2989}

\bibitem[\protect\citeauthoryear{{Fialkov} \& {Loeb}}{{Fialkov} \&
  {Loeb}}{2016}]{Fialkov_and_Loeb2016_EoR}
{Fialkov} A.,  {Loeb} A.,  2016, \mn@doi [\jcap]
  {10.1088/1475-7516/2016/05/004}, \href
  {https://ui.adsabs.harvard.edu/abs/2016JCAP...05..004F} {2016, 004}

\bibitem[\protect\citeauthoryear{{Foreman-Mackey} et~al.,}{{Foreman-Mackey}
  et~al.}{2019}]{emcee_v3}
{Foreman-Mackey} D.,  et~al., 2019, \mn@doi [The Journal of Open Source
  Software] {10.21105/joss.01864}, \href
  {https://ui.adsabs.harvard.edu/abs/2019JOSS....4.1864F} {4, 1864}

\bibitem[\protect\citeauthoryear{{Gourdji}, {Michilli}, {Spitler}, {Hessels},
  {Seymour}, {Cordes}  \& {Chatterjee}}{{Gourdji}
  et~al.}{2019}]{Gourdji2019_121102_arecibo}
{Gourdji} K.,  {Michilli} D.,  {Spitler} L.~G.,  {Hessels} J.~W.~T.,  {Seymour}
  A.,  {Cordes} J.~M.,   {Chatterjee} S.,  2019, \mn@doi [\apjl]
  {10.3847/2041-8213/ab1f8a}, \href
  {https://ui.adsabs.harvard.edu/abs/2019ApJ...877L..19G} {877, L19}

\bibitem[\protect\citeauthoryear{{Gupta}, {Mathur}, {Krongold}, {Nicastro}  \&
  {Galeazzi}}{{Gupta} et~al.}{2012}]{Gupta_et_al_2012_MW_Halo_EM}
{Gupta} A.,  {Mathur} S.,  {Krongold} Y.,  {Nicastro} F.,   {Galeazzi} M.,
  2012, \mn@doi [\apjl] {10.1088/2041-8205/756/1/L8}, \href
  {https://ui.adsabs.harvard.edu/abs/2012ApJ...756L...8G} {756, L8}

\bibitem[\protect\citeauthoryear{{Gupta} et~al.,}{{Gupta}
  et~al.}{2019}]{GuptaAtel2019c_FRB191107}
{Gupta} V.,  et~al., 2019, The Astronomer's Telegram, \href
  {https://ui.adsabs.harvard.edu/abs/2019ATel13282....1G} {13282, 1}

\bibitem[\protect\citeauthoryear{{Gupta} et~al.,}{{Gupta}
  et~al.}{2021}]{Gupta_et_al_2021_mock_frb_injection_utmost}
{Gupta} V.,  et~al., 2021, \mn@doi [\mnras] {10.1093/mnras/staa3683}, \href
  {https://ui.adsabs.harvard.edu/abs/2021MNRAS.501.2316G} {501, 2316}

\bibitem[\protect\citeauthoryear{{Hashimoto} et~al.,}{{Hashimoto}
  et~al.}{2021}]{Hashimoto_et_al2021EoR}
{Hashimoto} T.,  et~al., 2021, \mn@doi [\mnras] {10.1093/mnras/stab186}, \href
  {https://ui.adsabs.harvard.edu/abs/2021MNRAS.502.2346H} {502, 2346}

\bibitem[\protect\citeauthoryear{{Heintz} et~al.,}{{Heintz}
  et~al.}{2020}]{Heintz_2020_host_galaxies_ASKAP}
{Heintz} K.~E.,  et~al., 2020, \mn@doi [\apj] {10.3847/1538-4357/abb6fb}, \href
  {https://ui.adsabs.harvard.edu/abs/2020ApJ...903..152H} {903, 152}

\bibitem[\protect\citeauthoryear{{Hotan}, {van Straten}  \&
  {Manchester}}{{Hotan} et~al.}{2004}]{PSRCHIVE2004}
{Hotan} A.~W.,  {van Straten} W.,   {Manchester} R.~N.,  2004, \mn@doi [\pasa]
  {10.1071/AS04022}, \href
  {https://ui.adsabs.harvard.edu/abs/2004PASA...21..302H} {21, 302}

\bibitem[\protect\citeauthoryear{{Hunter}}{{Hunter}}{2007}]{matplotlib}
{Hunter} J.~D.,  2007, \mn@doi [Computing in Science \& Engineering]
  {10.1109/MCSE.2007.55}, 9, 90

\bibitem[\protect\citeauthoryear{{Inoue}}{{Inoue}}{2004}]{Inoue}
{Inoue} S.,  2004, \mn@doi [\mnras] {10.1111/j.1365-2966.2004.07359.x}, \href
  {https://ui.adsabs.harvard.edu/abs/2004MNRAS.348..999I} {348, 999}

\bibitem[\protect\citeauthoryear{{Ioka}}{{Ioka}}{2003}]{Ioka2003}
{Ioka} K.,  2003, \mn@doi [\apjl] {10.1086/380598}, \href
  {https://ui.adsabs.harvard.edu/abs/2003ApJ...598L..79I} {598, L79}

\bibitem[\protect\citeauthoryear{Jeffreys}{Jeffreys}{1998}]{jeffreys1998theory_bayes_factor_citation}
Jeffreys H.,  1998, The theory of probability.
OUP Oxford

\bibitem[\protect\citeauthoryear{{Koay} et~al.,}{{Koay}
  et~al.}{2012}]{Koay2012_AGN_scattering_limit}
{Koay} J.~Y.,  et~al., 2012, \mn@doi [\apj] {10.1088/0004-637X/756/1/29}, \href
  {https://ui.adsabs.harvard.edu/abs/2012ApJ...756...29K} {756, 29}

\bibitem[\protect\citeauthoryear{{Lee} \& {Jokipii}}{{Lee} \&
  {Jokipii}}{1975}]{Lee_and_Jokipii_1975_scattering_theory}
{Lee} L.~C.,  {Jokipii} J.~R.,  1975, \mn@doi [\apj] {10.1086/153916}, \href
  {https://ui.adsabs.harvard.edu/abs/1975ApJ...201..532L} {201, 532}

\bibitem[\protect\citeauthoryear{{Lee} \& {Jokipii}}{{Lee} \&
  {Jokipii}}{1976}]{Lee_and_Jokiipi1976}
{Lee} L.~C.,  {Jokipii} J.~R.,  1976, \mn@doi [\apj] {10.1086/154434}, \href
  {https://ui.adsabs.harvard.edu/abs/1976ApJ...206..735L} {206, 735}

\bibitem[\protect\citeauthoryear{{Lorimer} \& {Kramer}}{{Lorimer} \&
  {Kramer}}{2004}]{Pulsar_Handbook_2004_Lorimer_and_Kramer}
{Lorimer} D.~R.,  {Kramer} M.,  2004, {Handbook of Pulsar Astronomy}.
~ Vol. 4

\bibitem[\protect\citeauthoryear{{Lorimer}, {Bailes}, {McLaughlin}, {Narkevic}
  \& {Crawford}}{{Lorimer} et~al.}{2007}]{Lorimer2007}
{Lorimer} D.~R.,  {Bailes} M.,  {McLaughlin} M.~A.,  {Narkevic} D.~J.,
  {Crawford} F.,  2007, \mn@doi [Science] {10.1126/science.1147532}, \href
  {https://ui.adsabs.harvard.edu/abs/2007Sci...318..777L} {318, 777}

\bibitem[\protect\citeauthoryear{{Macquart} \& {Koay}}{{Macquart} \&
  {Koay}}{2013}]{Macquart&Koay2013}
{Macquart} J.-P.,  {Koay} J.~Y.,  2013, \mn@doi [\apj]
  {10.1088/0004-637X/776/2/125}, \href
  {https://ui.adsabs.harvard.edu/abs/2013ApJ...776..125M} {776, 125}

\bibitem[\protect\citeauthoryear{{Macquart}, {Shannon}, {Bannister}, {James},
  {Ekers}  \& {Bunton}}{{Macquart} et~al.}{2019}]{Macquart2019_spec_idx}
{Macquart} J.~P.,  {Shannon} R.~M.,  {Bannister} K.~W.,  {James} C.~W.,
  {Ekers} R.~D.,   {Bunton} J.~D.,  2019, \mn@doi [\apjl]
  {10.3847/2041-8213/ab03d6}, \href
  {https://ui.adsabs.harvard.edu/abs/2019ApJ...872L..19M} {872, L19}

\bibitem[\protect\citeauthoryear{{Macquart} et~al.,}{{Macquart}
  et~al.}{2020}]{Macquart2020_IGM_Baryons_DM_z_relation}
{Macquart} J.~P.,  et~al., 2020, \mn@doi [\nat] {10.1038/s41586-020-2300-2},
  \href {https://ui.adsabs.harvard.edu/abs/2020Natur.581..391M} {581, 391}

\bibitem[\protect\citeauthoryear{{Mannings} et~al.,}{{Mannings}
  et~al.}{2020}]{Mannings_2020_host_galaxies_env}
{Mannings} A.~G.,  et~al., 2020, arXiv e-prints, \href
  {https://ui.adsabs.harvard.edu/abs/2020arXiv201211617M} {p. arXiv:2012.11617}

\bibitem[\protect\citeauthoryear{{Marcote} et~al.,}{{Marcote}
  et~al.}{2020}]{Marcote_2020_180916_localisation_Nature}
{Marcote} B.,  et~al., 2020, \mn@doi [\nat] {10.1038/s41586-019-1866-z}, \href
  {https://ui.adsabs.harvard.edu/abs/2020Natur.577..190M} {577, 190}

\bibitem[\protect\citeauthoryear{{Masui} et~al.,}{{Masui}
  et~al.}{2015}]{Masui2015}
{Masui} K.,  et~al., 2015, \mn@doi [\nat] {10.1038/nature15769}, \href
  {https://ui.adsabs.harvard.edu/abs/2015Natur.528..523M} {528, 523}

\bibitem[\protect\citeauthoryear{{Niino}}{{Niino}}{2020}]{Niino_2020_host_galaxy_DM}
{Niino} Y.,  2020, arXiv e-prints, \href
  {https://ui.adsabs.harvard.edu/abs/2020arXiv200512891N} {p. arXiv:2005.12891}

\bibitem[\protect\citeauthoryear{{Nimmo} et~al.,}{{Nimmo}
  et~al.}{2021a}]{Nimmo2021_nanosecond_FRB200120E}
{Nimmo} K.,  et~al., 2021a, arXiv e-prints, \href
  {https://ui.adsabs.harvard.edu/abs/2021arXiv210511446N} {p. arXiv:2105.11446}

\bibitem[\protect\citeauthoryear{{Nimmo} et~al.,}{{Nimmo}
  et~al.}{2021b}]{Nimmo2021_microstructure_180916}
{Nimmo} K.,  et~al., 2021b, \mn@doi [Nature Astronomy]
  {10.1038/s41550-021-01321-3}, \href
  {https://ui.adsabs.harvard.edu/abs/2021NatAs...5..594N} {5, 594}

\bibitem[\protect\citeauthoryear{{Ocker}, {Cordes}  \& {Chatterjee}}{{Ocker}
  et~al.}{2021}]{Ocker_and_Cordes_2021_Halo_scattering_limits}
{Ocker} S.~K.,  {Cordes} J.~M.,   {Chatterjee} S.,  2021, \mn@doi [\apj]
  {10.3847/1538-4357/abeb6e}, \href
  {https://ui.adsabs.harvard.edu/abs/2021ApJ...911..102O} {911, 102}

\bibitem[\protect\citeauthoryear{{Oswald} et~al.,}{{Oswald}
  et~al.}{2021}]{Oswald21_Meerkat_pulsar_scattering_indices}
{Oswald} L.~S.,  et~al., 2021, \mn@doi [\mnras] {10.1093/mnras/stab980}, \href
  {https://ui.adsabs.harvard.edu/abs/2021MNRAS.504.1115O} {504, 1115}

\bibitem[\protect\citeauthoryear{{Pagano} \& {Fronenberg}}{{Pagano} \&
  {Fronenberg}}{2021}]{Pagano_et_al2021_EoR}
{Pagano} M.,  {Fronenberg} H.,  2021, \mn@doi [\mnras]
  {10.1093/mnras/stab1438}, \href
  {https://ui.adsabs.harvard.edu/abs/2021MNRAS.505.2195P} {505, 2195}

\bibitem[\protect\citeauthoryear{{Platts}, {Prochaska}  \& {Law}}{{Platts}
  et~al.}{2020}]{Platts2020_MW_halo}
{Platts} E.,  {Prochaska} J.~X.,   {Law} C.~J.,  2020, \mn@doi [\apjl]
  {10.3847/2041-8213/ab930a}, \href
  {https://ui.adsabs.harvard.edu/abs/2020ApJ...895L..49P} {895, L49}

\bibitem[\protect\citeauthoryear{{Prochaska} \& {Neeleman}}{{Prochaska} \&
  {Neeleman}}{2018}]{Prochaska_Neeleman_2018_DLAs}
{Prochaska} J.~X.,  {Neeleman} M.,  2018, \mn@doi [\mnras]
  {10.1093/mnras/stx2824}, \href
  {https://ui.adsabs.harvard.edu/abs/2018MNRAS.474..318P} {474, 318}

\bibitem[\protect\citeauthoryear{{Prochaska} \& {Zheng}}{{Prochaska} \&
  {Zheng}}{2019}]{Prochaska2019_haloes}
{Prochaska} J.~X.,  {Zheng} Y.,  2019, \mn@doi [\mnras] {10.1093/mnras/stz261},
  \href {https://ui.adsabs.harvard.edu/abs/2019MNRAS.485..648P} {485, 648}

\bibitem[\protect\citeauthoryear{{Prochaska} et~al.,}{{Prochaska}
  et~al.}{2019}]{Prochaska2019_FRB181112}
{Prochaska} J.~X.,  et~al., 2019, arXiv e-prints, \href
  {https://ui.adsabs.harvard.edu/abs/2019arXiv190911681P} {p. arXiv:1909.11681}

\bibitem[\protect\citeauthoryear{{Qiu} et~al.,}{{Qiu}
  et~al.}{2020}]{Hao_qiu_2020_ASKAP_FRB_scattering_props}
{Qiu} H.,  et~al., 2020, \mn@doi [\mnras] {10.1093/mnras/staa1916}, \href
  {https://ui.adsabs.harvard.edu/abs/2020MNRAS.497.1382Q} {497, 1382}

\bibitem[\protect\citeauthoryear{{Ravi}}{{Ravi}}{2019}]{Ravi2019_observed_prop}
{Ravi} V.,  2019, \mn@doi [\mnras] {10.1093/mnras/sty1551}, \href
  {https://ui.adsabs.harvard.edu/abs/2019MNRAS.482.1966R} {482, 1966}

\bibitem[\protect\citeauthoryear{{Ravi} et~al.,}{{Ravi}
  et~al.}{2019a}]{Ravi2019_localisation}
{Ravi} V.,  et~al., 2019a, arXiv e-prints, \href
  {https://ui.adsabs.harvard.edu/abs/2019arXiv190701542R} {p. arXiv:1907.01542}

\bibitem[\protect\citeauthoryear{{Ravi} et~al.,}{{Ravi}
  et~al.}{2019b}]{Ravi_et_al2019_tomography_of_universe}
{Ravi} V.,  et~al., 2019b, \baas, \href
  {https://ui.adsabs.harvard.edu/abs/2019BAAS...51c.420R} {51, 420}

\bibitem[\protect\citeauthoryear{{Shannon} \& {Cordes}}{{Shannon} \&
  {Cordes}}{2010}]{Shannon_and_Cordes_2010_spin_noise_modelling}
{Shannon} R.~M.,  {Cordes} J.~M.,  2010, \mn@doi [\apj]
  {10.1088/0004-637X/725/2/1607}, \href
  {https://ui.adsabs.harvard.edu/abs/2010ApJ...725.1607S} {725, 1607}

\bibitem[\protect\citeauthoryear{{Shannon} et~al.,}{{Shannon}
  et~al.}{2018}]{Shannon2018}
{Shannon} R.~M.,  et~al., 2018, \mn@doi [\nat] {10.1038/s41586-018-0588-y},
  \href {https://ui.adsabs.harvard.edu/abs/2018Natur.562..386S} {562, 386}

\bibitem[\protect\citeauthoryear{{Simha} et~al.,}{{Simha}
  et~al.}{2020}]{Simha2020_190608_IGM_cosmic_web}
{Simha} S.,  et~al., 2020, \mn@doi [\apj] {10.3847/1538-4357/abafc3}, \href
  {https://ui.adsabs.harvard.edu/abs/2020ApJ...901..134S} {901, 134}

\bibitem[\protect\citeauthoryear{Speagle}{Speagle}{2020}]{dynesty_sampler2020}
Speagle J.~S.,  2020, \mn@doi [Monthly Notices of the Royal Astronomical
  Society] {10.1093/mnras/staa278}, 493, 3132–3158

\bibitem[\protect\citeauthoryear{{Spitler} et~al.,}{{Spitler}
  et~al.}{2016}]{Spitler2016}
{Spitler} L.~G.,  et~al., 2016, \mn@doi [\nat] {10.1038/nature17168}, \href
  {https://ui.adsabs.harvard.edu/abs/2016Natur.531..202S} {531, 202}

\bibitem[\protect\citeauthoryear{{The CHIME/FRB Collaboration} et~al.,}{{The
  CHIME/FRB Collaboration} et~al.}{2020}]{CHIME2020_SGR_burst}
{The CHIME/FRB Collaboration} et~al., 2020, arXiv e-prints, \href
  {https://ui.adsabs.harvard.edu/abs/2020arXiv200510324T} {p. arXiv:2005.10324}

\bibitem[\protect\citeauthoryear{{The CHIME/FRB Collaboration} et~al.,}{{The
  CHIME/FRB Collaboration} et~al.}{2021}]{First_CHIME_catalog_2021}
{The CHIME/FRB Collaboration} et~al., 2021, arXiv e-prints, \href
  {https://ui.adsabs.harvard.edu/abs/2021arXiv210604352T} {p. arXiv:2106.04352}

\bibitem[\protect\citeauthoryear{Trotta}{Trotta}{2008}]{Trotta_2008_Bayes_factor_Jeffreys_scale}
Trotta R.,  2008, \mn@doi [Contemporary Physics] {10.1080/00107510802066753},
  49, 71

\bibitem[\protect\citeauthoryear{{Vedantham} \& {Phinney}}{{Vedantham} \&
  {Phinney}}{2019}]{Vedantham2019_scattering_CGM}
{Vedantham} H.~K.,  {Phinney} E.~S.,  2019, \mn@doi [\mnras]
  {10.1093/mnras/sty2948}, \href
  {https://ui.adsabs.harvard.edu/abs/2019MNRAS.483..971V} {483, 971}

\bibitem[\protect\citeauthoryear{{W}es {M}c{K}inney}{{W}es
  {M}c{K}inney}{2010}]{pandas-official}
{W}es {M}c{K}inney 2010, in {S}t\'efan van~der {W}alt {J}arrod {M}illman eds,
  {P}roceedings of the 9th {P}ython in {S}cience {C}onference. pp 56 -- 61,
  \mn@doi{10.25080/Majora-92bf1922-00a}

\bibitem[\protect\citeauthoryear{Xu \& Zhang}{Xu \&
  Zhang}{2016}]{Xu_and_Zhang_2016_origin_of_scattering}
Xu S.,  Zhang B.,  2016, \mn@doi [The Astrophysical Journal]
  {10.3847/0004-637x/832/2/199}, 832, 199

\bibitem[\protect\citeauthoryear{{Yao}, {Manchester}  \& {Wang}}{{Yao}
  et~al.}{2017}]{YMW16}
{Yao} J.~M.,  {Manchester} R.~N.,   {Wang} N.,  2017, \mn@doi [\mnras]
  {10.1093/mnras/stx729}, \href
  {https://ui.adsabs.harvard.edu/abs/2017MNRAS.468.3289Y} {468, 3289}

\bibitem[\protect\citeauthoryear{{Zhang}}{{Zhang}}{2018}]{BingZhang2018_detectibility_of_high_z_FRBs}
{Zhang} B.,  2018, \mn@doi [\apjl] {10.3847/2041-8213/aae8e3}, \href
  {https://ui.adsabs.harvard.edu/abs/2018ApJ...867L..21Z} {867, L21}

\bibitem[\protect\citeauthoryear{{Zhu}, {Feng}  \& {Zhang}}{{Zhu}
  et~al.}{2018}]{Zhu_Feng_2018_scattering_hydrosims}
{Zhu} W.,  {Feng} L.-L.,   {Zhang} F.,  2018, \mn@doi [\apj]
  {10.3847/1538-4357/aadbb0}, \href
  {https://ui.adsabs.harvard.edu/abs/2018ApJ...865..147Z} {865, 147}

\bibitem[\protect\citeauthoryear{{van der Walt}, {Colbert}  \&
  {Varoquaux}}{{van der Walt} et~al.}{2011}]{numpy}
{van der Walt} S.,  {Colbert} S.~C.,   {Varoquaux} G.,  2011, \mn@doi
  [Computing in Science and Engineering] {10.1109/MCSE.2011.37}, \href
  {https://ui.adsabs.harvard.edu/abs/2011CSE....13b..22V} {13, 22}

\makeatother
\end{thebibliography}








\bsp	
\label{lastpage}
\end{document}